%% file: TOP-11-021_temp.tex
\pdfoutput=1

\documentclass[11pt,twoside,a4paper,cmspaper,final,collab]{cms-tdr}
\def\svnVersion{177020}\def\cmsCernNo{2012-274}\def\cmsCernDate{\today}\def\cmsMessage{Submitted to the Journal of High Energy Physics}

\begin{document}\cmsNoteHeader{TOP-11-021}

\hyphenation{had-ron-i-za-tion}
\hyphenation{cal-or-i-me-ter}
\hyphenation{de-vices}

\RCS$Revision: 156911 $
\RCS$HeadURL: svn+ssh://svn.cern.ch/reps/tdr2/papers/TOP-11-021/trunk/TOP-11-021.tex $
\RCS$Id: TOP-11-021.tex 156911 2012-11-07 17:55:38Z alverson $

\newcommand{\mt}{\ensuremath{m_{\ell\nu \cPqb}}\xspace}
\newcommand{\cosThetaPol}{\ensuremath{\cos{\theta^*_{lj}}}\xspace}
\newcommand{\etalj}{\ensuremath{|\eta_{j^\prime}|}\xspace}
\newcommand{\ppbar}{\Pp\Pap\xspace}
\renewcommand{\ttbar}{\cPqt\cPaqt\xspace}
\newcommand{\sigmatch}{\ensuremath{\sigma_{t\text{-ch.}}}\xspace}
\newcommand{\met}{\ensuremath{{E\!\!\!/}_{\!\mathrm{T}}}\xspace}
\renewcommand{\MET}{\met}
\newcommand{\mT}{\ensuremath{m_{\mathrm{T}}}\xspace}
\newcommand{\mTW}{\mT}
\newcommand{\PTslashsmall}{\ensuremath{{p}_\mathrm{T}\hspace{-0.77em}/\kern 0.5em}\xspace}
\newcommand{\midrule}{\hline}
\newcommand{\lumiMu}{1170\xspace}
\newcommand{\lumiEle}{1563\xspace}
\newcommand{\lumiMuFB}{1.17\xspace}
\newcommand{\lumiEleFB}{1.56\xspace}
\newcommand{\lumiEleSingleEle}{216\xspace}
\newcommand{\lumiEleCrossTrig}{1347\xspace}
\newlength\cmsFigWidth\setlength\cmsFigWidth{0.49\linewidth}

\cmsNoteHeader{TOP-11-021} 

\title{Measurement of the single-top-quark $t$-channel cross section in pp collisions at $\sqrt{s}=7$\TeV}

\date{\today}

\abstract{
A measurement of the single-top-quark $t$-channel production cross section in pp collisions at $\sqrt{s}=7$\TeV with the CMS detector at the LHC is presented.
Two different and complementary approaches have been followed.
The first approach exploits the distributions of the pseudorapidity of the recoil jet and reconstructed top-quark mass using
background estimates determined from control samples in data.
The second approach is based on multivariate analysis techniques that probe the compatibility of the candidate events with the signal.
Data have been collected for the muon and electron final states, corresponding to integrated luminosities of \lumiMuFB and \lumiEleFB\fbinv, respectively. The single-top-quark production cross section in the $t$-channel is measured to be
$67.2\pm 6.1\text{pb}$, in agreement with the approximate next-to-next-to-leading-order standard model prediction. Using the standard model electroweak couplings, the CKM matrix element $|V_{\text{tb}}|$ is measured to be $1.020\pm 0.046\text{ (meas.)} \pm0.017\text{ (theor.)}$.
}
\hypersetup{%
pdfauthor={t-channel single top subgroup},%
pdftitle={Measurement of the single-top-quark t-channel cross section in pp collisions at sqrt(s) = 7 TeV},%
pdfsubject={CMS},%
pdfkeywords={CMS, physics, top, quark, topquark, top-quark, single-top-quark, singletop, single-top, BDT, NN, boosted decision trees, neural network}}

\maketitle 

\section{Introduction}
\label{sec:intro}
Single top quarks can be produced through charged-current electroweak interactions. Due to the
large top-quark mass, these processes are well suited to test the predictions of the standard model (SM) of particle
physics and to search for new phenomena. Measurements of the single-top-quark production cross section also provide an unbiased determination of the magnitude of
the Cabibbo--Kobayashi--Maskawa (CKM) matrix element $|V_{\mathrm{tb}}|$.

Single-top-quark production was observed in proton--antiproton collisions at the Tevatron collider with a centre-of-mass energy of $1.96$\TeV~\cite{CDF-singletop,D0-singletop, Group:2009qk}. The cross section increases by a factor of $20$ at the Large Hadron Collider (LHC) with respect to the Tevatron. The first measurements of the single-top-quark production cross section in proton--proton collisions at a centre-of-mass energy of $7$\TeV were performed by the Compact Muon Solenoid (CMS)~\cite{Chatrchyan:2011vp} and ATLAS~\cite{Aad:2012ux, :2012dj} experiments.

Previous measurements are compatible with expectations based on approximate next-to-leading-order and next-to-next-to-leading logarithm (NLO+NNLL) perturbative quantum chromodynamics (QCD) calculations.
In these calculations, three types of parton scattering processes are considered: $t$-channel and $s$-channel processes, and W-associated single-top-quark production (tW). The dominant contribution to the cross section is expected to be from the $t$-channel process with a cross section of
$\sigmatch^\mathrm{th} =64.6^{+2.1}_{-0.7}\,^{+1.5}_{-1.7}\unit{pb}$~\cite{Kidonakis:2011wy} for a top-quark mass of
$m_\mathrm{t} = 172.5\GeVcc$.

\begin{figure}[h]
  \begin{center}
    \includegraphics[width=0.3\textwidth]{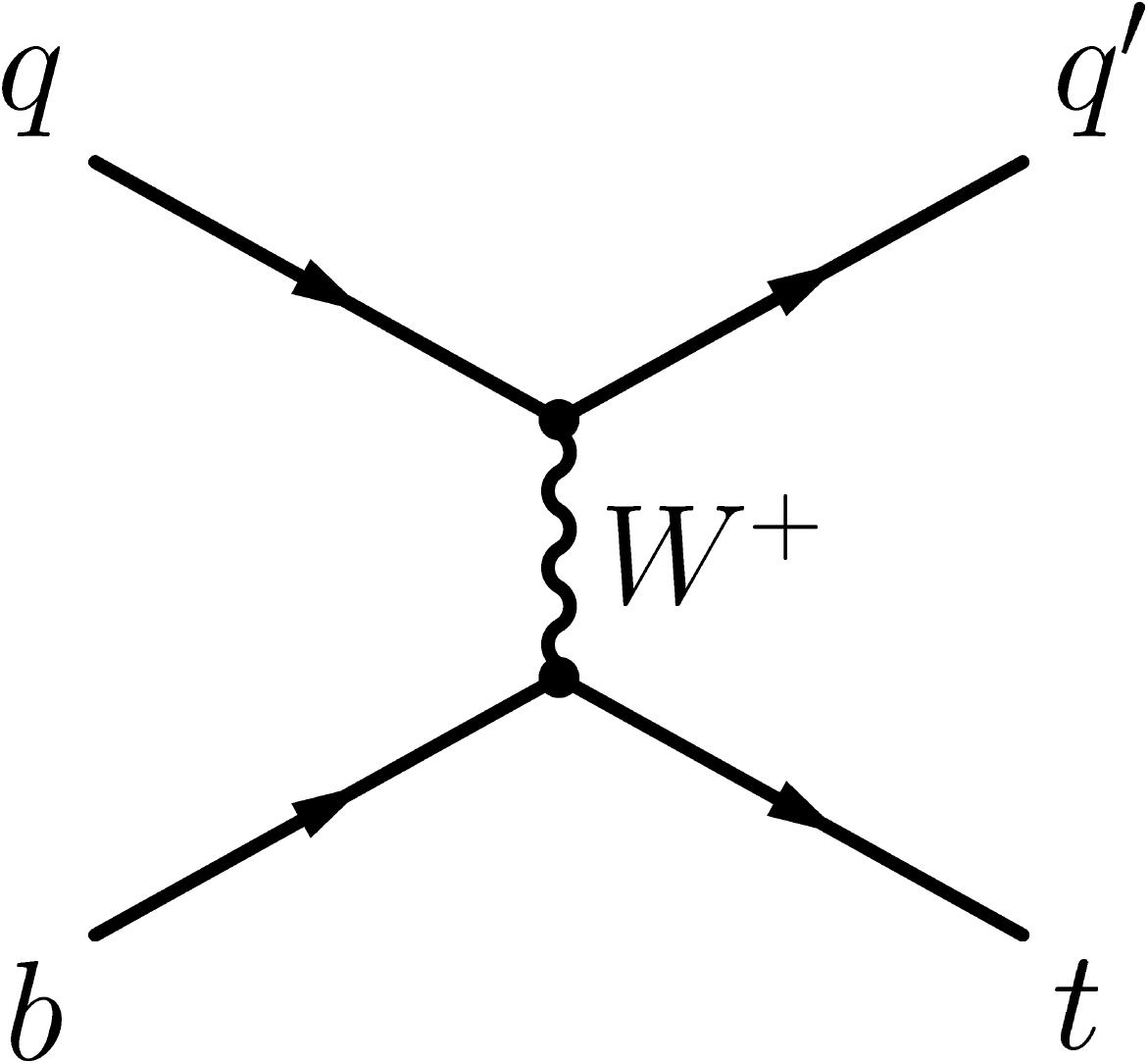}
    \includegraphics[width=0.3\textwidth]{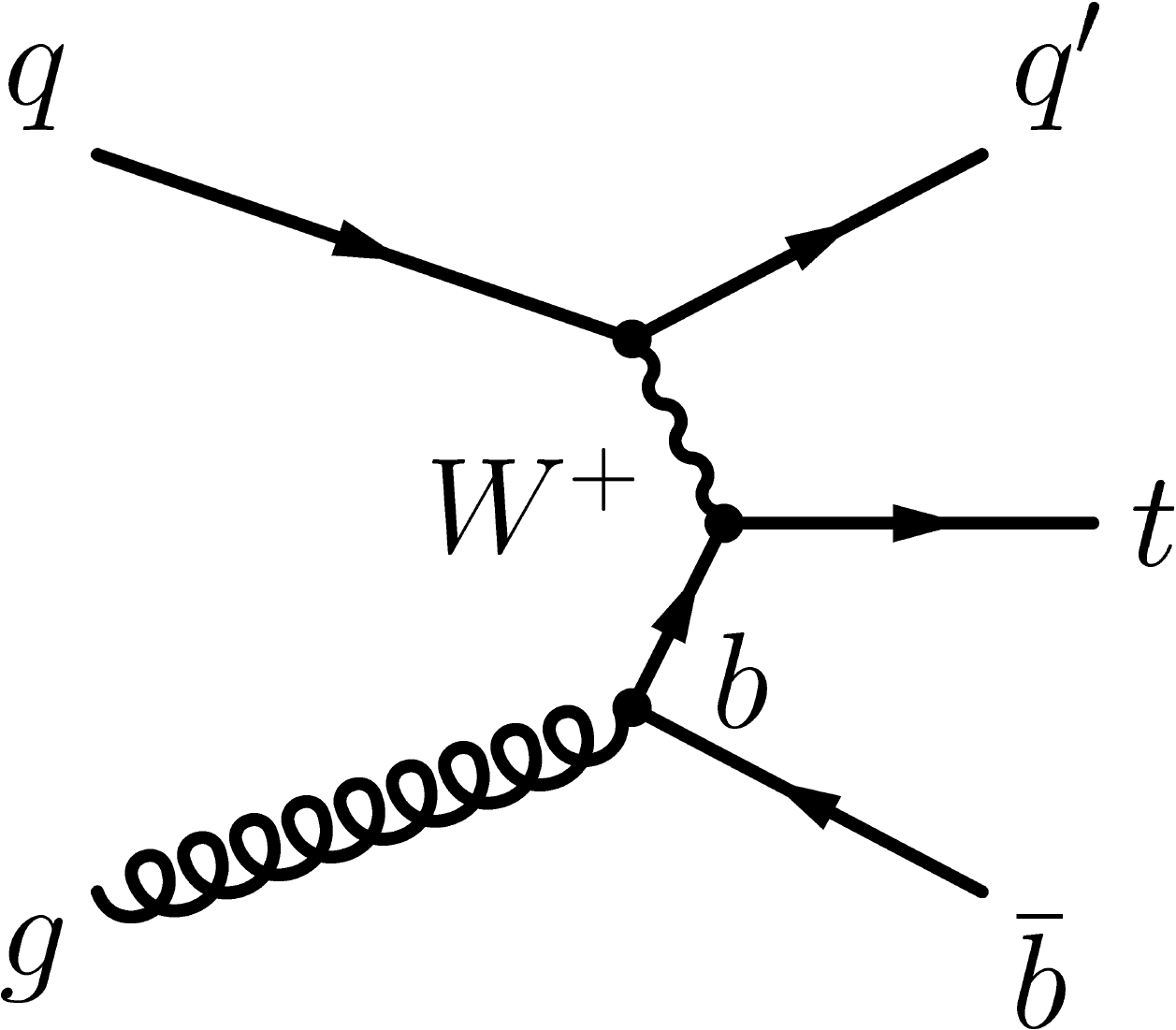}
    \caption{\label{fig:FG}Leading order Feynman diagrams for single-top-quark production in the $t$-channel: $2\to 2$ (left) and $2\to 3$ (right).}
  \end{center}
\end{figure}

This paper extends the previous CMS measurement~\cite{Chatrchyan:2011vp} of the $t$-channel cross section. The single-top-quark production cross section measurement is based on pp collision data at $\sqrt{s}=7$\TeV collected during 2011 with the CMS experiment, corresponding to integrated luminosities of \lumiMuFB and \lumiEleFB\fbinv with muon and electron final states, respectively.
Events with leptonically decaying W bosons are selected: $\mathrm{t}\to \mathrm{bW }\to \mathrm{b}\ell\nu$ ($\ell=\mathrm{e}$ or $\mu$). This measurement is used to determine the CKM matrix element $|V_{\mathrm{tb}}|$.

The $t$-channel event signature (Fig.~\ref{fig:FG}) typically comprises one forward jet
scattered off a top quark. The decay products of the top quark mainly appear in the central region of the detector.
A dedicated event selection is applied, and then measurements with two complementary approaches
are performed.
The first approach exploits the reconstructed top-quark mass and one of the angular properties
specific to $t$-channel top-quark production: the forward pseudorapidity
distribution of the light jet recoiling against the top quark. This analysis is referred to as the $\etalj$ analysis. It is straightforward and robust and has little model dependence.
The second approach exploits, via multivariate discriminators, the compatibility of the
signal candidates with the event characteristics predicted by the SM
for electroweak top-quark production. This approach aims for a precise $t$-channel cross section
measurement by optimising the discrimination between signal and background. The systematic uncertainties are constrained by simultaneously analysing phase space regions
with substantial $t$-channel single-top-quark contributions and regions where they are negligible.
Because of the higher complexity of this approach, two independent multivariate analyses are
conducted that cross-check each other. One is based on a Neural Network (NN) and the other on Boosted Decision Trees (BDT).
After validating the consistency of the results, the final result is determined by combining the three analyses using the Best Linear Unbiased Estimator (BLUE)~\cite{BLUE}.

\section{CMS Detector}
\label{sec:cms}

The CMS Detector is described in detail elsewhere~\cite{JINST}, and the key components for this analysis are listed below. The main feature of the CMS detector is a superconducting solenoidal magnet with a length of 13\unit{m} and
a diameter of 6~m which provides an axial magnetic field of 3.8\unit{T}.  The bore of the solenoid contains various particle detection systems.
Charged particle trajectories are recorded by the silicon pixel and strip tracker covering the region $|\eta|<2.5$, where the pseudorapidity $\eta$
is defined as $\eta=-\ln[\tan(\theta/2)]$, with $\theta$ the polar angle of the trajectory of the particle with respect to the anticlockwise
beam direction.  A crystal electromagnetic calorimeter (ECAL) and a brass/scintillator hadronic calorimeter (HCAL) surround the tracking volume and
extend to $|\eta|<3.0$. A quartz-fibre and steel absorber Cherenkov-light hadron forward detector extends the calorimeter coverage to $|\eta|<5.0$.
 Muons are detected in gas-ionization detectors embedded in the steel return yoke outside the solenoid.
 The detector is nearly hermetic, which permits good energy balance measurements in the plane transverse to the beam line.

\section{Collision Data and Simulation}
\label{sec:datasim}

The present analyses use data sets collected during 2011 corresponding to integrated luminosities of \lumiMuFB
and \lumiEleFB\fbinv for the muon and electron channels, respectively.

Single-top-quark $t$-channel signal events from Monte Carlo (MC) simulation used in this study have been generated with the next-to-leading-order (NLO) MC generator
{\sc powheg}~\cite{Re:2010bp,Alioli:2010xd,Alioli:2009je,Frixione:2007vw} interfaced to \PYTHIA~6.4.24~\cite{pythia} for the parton showering simulation. This
ensures good modelling of the 2$\rightarrow$2 (Fig.~\ref{fig:FG}, left) process and all NLO contributions, in particular of the 2$\rightarrow$3
diagram (Fig.~\ref{fig:FG}, right). Signal events have also been generated using \COMPHEP~\cite{comphep} to study systematic uncertainties related to the MC generator. A top-quark mass of $172.5\GeVcc$ has been assumed in the simulations.

Several SM processes are taken into account as background to the analysis. The \POWHEG generator interfaced with \PYTHIA is used
to model the single-top-quark tW and $s$-channel events, which are considered as background in this analysis.
The tree-level matrix-element generator \MADGRAPH~\cite{madgraph} interfaced to \PYTHIA
 is used for top-quark pair (\ttbar) production and for inclusive single boson production
V + X, where V = W or Z, and X can indicate one or more
light or heavy partons. In the following, we will explicitly distinguish the processes with V+light partons, Wc, and VQ$\overline{\mathrm{Q}}$, with Q = b or c.
A procedure implemented during the event generation and based on the ``MLM prescription''~\cite{Alwall:2007fs} avoids double counting between matrix element and parton shower generated jets.
The remaining background samples are simulated using \PYTHIA. These include diboson production (WW, WZ, ZZ), $\gamma+$jets, and multijet
QCD enriched events with electrons or muons coming from the decays of b and c quarks, as well as muons from the decay of long-lived hadrons.
The \textsc{cteq6}~\cite{PDF:CTEQ6} parton distribution functions are used for all simulated samples.

All generated events undergo a full simulation of the detector response according to the CMS implementation of \GEANTfour~\cite{geant},
and are processed by the same reconstruction software used for collision data.
\section{Event Selection and Reconstruction}
\label{sec:selection}

Events are characterized by a single isolated muon or electron and momentum imbalance due to the presence of a neutrino, with one central b jet from the top-quark decay.
An additional light-quark jet from the hard-scattering process is often present in the
forward direction. A second b jet produced in association with the top quark can also be present (Fig.~\ref{fig:FG}, right),
although it yields a softer $\pt$ spectrum with respect to the b jet coming from the top-quark decay.

The trigger used for the online selection of the analysed data for the muon channel is based
on the presence of at least one isolated muon with a
transverse momentum $\pt>17\GeVc$.
For the electron channel, an isolated electron trigger with a transverse momentum $\pt>$~27\GeVc was used for the initial data-taking period, corresponding to an integrated luminosity of \lumiEleSingleEle~\pbinv. For the remaining data-taking period, a trigger selecting at least one electron with $\pt>25\GeVc$ and a jet with $\pt>30\GeVc$ was used. The jet is identified in the trigger processing as coming from a fragmentation of a b quark using the Track Counting High-Efficiency (TCHE) b-tagging algorithm described in Ref.~\cite{btag}.

At least one primary vertex is required, reconstructed from a minimum of four tracks, with a longitudinal distance of less than 24\unit{cm}, and a radial distance of less than 2\unit{cm} from the nominal interaction point. To select good muon and electron candidates, the same lepton identification as described in Ref.~\cite{top-11-014} is applied. Electrons,
muons, photons, and charged and neutral hadron candidates are reconstructed and identified using the CMS particle-flow (PF) algorithm~\cite{pflow}. The missing transverse momentum vector $\vec{\PTslash}$ is reconstructed from the momentum imbalance of PF particle candidates in the plane transverse to the beams direction. The magnitude of $\vec{\PTslash}$ is the missing transverse energy $\MET$.
The presence of exactly one isolated muon or electron candidate originating from the primary vertex is required in the event.
Muon candidates are selected by requiring a transverse momentum $\pt>20\GeVc$ and a pseudorapidity $|\eta|<2.1$. Electron candidates must have a transverse momentum $\pt>30$\GeVc and $|\eta|<2.5$.
Lepton isolation $I_{\text{rel}}^{\ell}$ is defined as the sum of the transverse energy deposited by stable
charged hadrons, neutral hadrons, and photons in a cone of $\Delta R = \sqrt{(\Delta\eta)^2+(\Delta\phi)^2} = 0.4$ around the charged-lepton track,
divided by the transverse momentum of the lepton. Muon isolation is ensured by requiring $I_{\text{rel}}^{\ell}<0.15$,
while the isolation requirement is tightened to 0.125 for electrons.
An electron candidate is rejected if it is identified as originating from the
conversion of a photon into an electron--positron pair or if it fails the identification criteria described in Ref.~\cite{top-11-014}.
Events are also rejected if an additional muon candidate is present that passes looser quality criteria, namely, $\pt >$ 10\GeVc, $|\eta| < 2.5$, and $I_{\text{rel}}^{\ell}<$ 0.2. For additional electrons, the required transverse momentum is $\pt >15\GeVc$.

Jets are defined by clustering PF candidates according to the anti-\kt algorithm~\cite{Cacciari:2008gp} with a distance parameter of
0.5. The analysis considers jets whose calibrated transverse momentum is greater than 30\GeVc for  $|\eta|<4.5$.
An event is accepted for further analysis only if at least two such jets are reconstructed. A jet is identified as coming from a b-quark fragmentation if it passes a
tight threshold on the Track Counting High-Purity (TCHP) b-tagging algorithm~\cite{btag} corresponding to a misidentification probability of 0.1\%.
The difference between simulated and measured b-tagging efficiencies, for true and misidentified b jets, is corrected by scaling the simulated event yields according to $\pt$-dependent scale factors determined from control samples from the data~\cite{btag}.

Events with a muon that is not from a leptonic decay of a W boson are suppressed by requiring a reconstructed transverse W boson mass
$\mT= \sqrt{2\pt\MET(1-\cos(\Delta\phi_{\ell,\vec{\PTslashsmall}}))}>40$\GeVcc, where $\Delta\phi_{\ell,\vec{\PTslashsmall}}$
is the azimuthal angle between the muon and the $\vec{\PTslash}$ directions and \pt is the transverse momentum of the muon.
For the electron channel, where the QCD multijet contamination is larger, the requirement $\MET >35\GeV$ is applied instead of the $\mT$ selection.

To classify signal and control samples, different event categories are defined and
denoted as ``$n$-jets $m$-btags'', where $n$ is the number of selected jets (2, 3, or 4) and $m$
is the number of selected b-tagged jets (0, 1, or $\geq$ 2).
The single-top-quark $t$-channel signal is primarily contained in the category ``2-jets 1-btag'', followed by ``3-jets 1-btag'', as the second b jet, which is produced in association with the top quark, is mostly out of acceptance.
The other categories are dominated by background processes with different compositions.
In particular, the ``2/3-jets 0-btags'' categories are enriched in events with a W boson produced in association with light partons (u, d, s, g). The ``3-jets 2-btags'' and ``4-jets 0/1/2-btags'' categories are enriched in $\ttbar$ events.

To extract the signal content, the NN and BDT analyses utilize the following six categories simultaneously for the measurement of the signal cross section: ``2-jets 1-btag'', ``3-jets 1-btag'', ``4-jets 1-btag'', ``2-jets 2-btags'',
``3-jets 2-btags'', and ``4-jets 2-btags''.
The latter four categories are used to constrain nuisance parameters (\eg the b-tagging efficiency or background normalisation). Besides these six categories, the three 0-tag categories are used to check the modelling of input variables.
The $\etalj$ analysis extracts the signal content from the ``2-jets 1-btag'' category only,
but uses the ``2-jets 0-btags'' and ``3-jets 2-btags'' categories to check the modelling of backgrounds.

The reconstruction of the top quark from its decay products leads to multiple
choices of possible top-quark candidates.
In the first step, the W-boson candidate is reconstructed from the charged lepton and from $\vec{\PTslash}$ following the procedure described in Ref.~\cite{Chatrchyan:2011vp}.
In the second step, the top-quark candidate is reconstructed by combining the W-boson candidate
with a jet identified as coming from a b quark, and its mass $\mt$ is calculated. Depending on the analysis category, the
ambiguity in the choice of the b-quark jet from the top-quark decay and the recoiling light
quark has to be resolved.
Events in the ``2-jets 1-btag'' category have no ambiguity: the b-tagged jet is associated with the top-quark decay, and the other jet is considered to be a light-quark jet.

In the other categories, the top quark and light-quark jet reconstruction has been optimized
for the purpose of each analysis and differs among them.
In the NN and BDT analyses, the most forward jet is chosen to be the light-quark jet in categories
where two jets and zero or two b tags are required. The other jet is associated with the
top-quark decay. In categories where three or more jets are required, in the case of one or
two required b tags, the jet with the lowest value of the TCHP discriminator is assumed
to originate from the light quark. If no b tag is required, the most forward jet is associated
with the light quark. From the remaining jets, the one which together with the reconstructed W boson
has a reconstructed top-quark mass $\mt$ closest to $m_{\mathrm{t}}=172.5\GeVcc$
is chosen as the b jet coming from the top-quark decay.
In the $\etalj$ analysis, the jet with the highest value of the TCHP discriminator is used for the
top-quark reconstruction in the ``2-jets 0-btags'' and ``3-jets 2-btags'' categories. The inclusive
$|\eta|$ distribution of both jets is used in ``2-jets 0-btags'', while for the ``3-jets 2-btags''
category the $|\eta|$ of the non-b-tagged jet is used.

In the $\etalj$ analysis the invariant mass $\mt$ of the reconstructed top quark is used to
further divide the ``2-jets 1-btag'' category into a $t$-channel enriched signal region (SR), defined
by selecting events within the mass range $130< \mt <220$\GeVcc, and a W boson and
$\ttbar$ enriched sideband region (SB), defined by selecting events that are outside this $\mt$ mass
window. The event yield in the SR is summarised in Table~\ref{tab:yield} for the muon and electron channels, together with expectations from
simulated signal and backgrounds, and for QCD multijet events, which are determined from control samples of data, as described in Section~\ref{sec:qcd}.

\begin{table}[htp]
\topcaption{Event yield with statistical uncertainties of the $\etalj$ analysis for the signal and main background processes in the signal region, after applying the $\mt$ mass requirement for the $\mu$ and e channels.
The yields are taken from simulation except for the QCD multijet yield, which is obtained from control samples of data as described in Section~\ref{sec:qcd}. The normalisation of the Wc($\bar{\mathrm{c}}$) and Wb($\bar{\mathrm{b}}$) processes is further discussed in Section~\ref{sec:whfextraction}.}
\label{tab:yield}
\begin{center}
  \begin{tabular}{ lcc }
\hline
Process & Muon yield & Electron yield \\
\hline\hline
$t$-channel & 617 $\pm$ 3 & 337 $\pm$ 2 \\
\hline
tW channel & 107 $\pm$ 1 & 70.2 $\pm$ 0.9 \\
$s$-channel & 25.6 $\pm$ 0.5 & 14.7 $\pm$ 0.4 \\
$\ttbar$ & 661 $\pm$ 6 & 484 $\pm$ 5 \\
W + light partons & 92 $\pm$ 7 & 38 $\pm$ 4 \\
Wc($\overline{\mathrm{c}}$) & 432 $\pm$ 14 & 201 $\pm$ 9 \\
Wb($\overline{\mathrm{b}}$) & 504 $\pm$ 14 & 236 $\pm$ 10 \\
Z + jets & 87 $\pm$ 3 & 13 $\pm$ 1 \\
Dibosons & 23.3 $\pm$ 0.4 & 10.7 $\pm$ 0.3 \\
QCD multijet& 77 $\pm$ 3 & 62 $\pm$ 3 \\
\hline
Total & 2626 $\pm$ 22 & 1468 $\pm$ 16 \\
\hline
Data & 3076 & 1588\\
\hline
\end{tabular}
\end{center}
\end{table}

\section{Background Estimation and Control Samples}
\label{sec:bkg}

Several control data samples  are used for two main purposes:
\begin{itemize}
\item to check the distributions of variables used as inputs to the analyses and the agreement between data and simulation;
\item to determine from data the yields and distributions of variables of interest for the main background processes.
\end{itemize}

All the analyses use rate and shape determinations from QCD multijet background data. The $\etalj$ analysis also determines
the yields and distributions of the background processes by using W boson production in association with jets from light quarks as well as c and b quarks. The NN and BDT analyses take the shape of the W+jets background from MC simulation, but consider the impact of shape uncertainties in the evaluation of the systematic uncertainties. In all three analyses the rate of the W+jets background is determined ``in situ'' by the signal extraction method as described in Section \ref{sec:bayes}.

\subsection{QCD Multijet Background Estimation}
\label{sec:qcd}

The yield of the QCD multijet background
in the different categories
is measured by performing fits to the distributions of the transverse W boson
mass $\mT$ in the muon channel and $\MET$ in the electron channel. A maximum-likelihood fit to the distribution of $\mT$ or $\MET$ is
performed assuming the following parameterisation: $F(x)= a\cdot S(x)+b\cdot B(x)$, where $x$ is $\mT$ or $\MET$ for the muon and electron
channels, respectively, and $S(x)$ and $B(x)$ are the expected distributions for the sum of all processes including a W boson in the final state and for
QCD multijet events, respectively. The function $S(x)$ is taken from simulation, while $B(x)$ is extracted directly from data.
The parameters $a$ and $b$ are determined from the fit.
The QCD multijet background yield is estimated to be the area of the fitted curve $b\cdot B(x)$
in the range  $\mT>40\GeVcc$ for the muon channel
and $\MET>35\GeV$ for the electron channel, as mentioned in Section~\ref{sec:selection}.

QCD multijet enriched samples from data are used to model the distributions $B(x)$ of $\mT$ and $\MET$.
For the muon channel, this sample is selected by inverting the isolation requirement. For the electron channel,
the selected electron is required to fail at least two of the three following quality requirements:
$I_{\text{rel}}^{\ell}<0.1$, the distance of closest approach to the primary vertex on the $x$-$y$ plane $\delta_{xy}<0.02$~cm,
and the electron identification criteria given in Ref.~\cite{top-11-014}.
It was verified by simulation that the $\mT$ and $\MET$ distributions for
QCD multijet like events are not significantly affected by this altered event
selection.

In the $\etalj$ analysis the fits to the $\mT$ and $\MET$ distributions cannot be performed reliably in the SR and SB separately due to the limited size of the simulated
samples, which would introduce large uncertainties in the signal modelling. For this reason, the fit is performed on the entire ``2-jets 1-btag'' sample. The number
of QCD multijet events in the SB and SR regions is determined by scaling the total QCD multijet yield, obtained from the fit, by the fraction of events in the two regions (SB and SR) of the $\mt$
distribution, as determined from the QCD multijet enriched region.
In the simulation, the distributions of the relevant variables obtained from the QCD multijet enriched sample are consistent with the ones in the SR.
The QCD multijet yields restricted to the SR are reported in Table~\ref{tab:yield}.

For all three analyses, the relative uncertainties on the QCD multijet yield estimates
are taken to be ${\pm}50\%$ for the muon channel and
${\pm}100\%$ for the electron channel.
Several cross checks have been performed; for example, the same fits have been
repeated taking $B(x)$ from simulation for both channels. Moreover, the choice of
$x=\mT$ or $\MET$ has been inverted for the muon and electron channels,
in this way performing fits on $\mT$ for the electron channel and $\MET$ for
the muon channel. The results in each case are in agreement with the previous
estimate within the assumed uncertainties.

\subsection{W+jets Background Estimation and Other Control Samples}
\label{sec:control}
\label{sec:whfextraction}

A check of the modelling for the W+jets background is carried out for each of the three analyes in the 0-tag control regions. In particular, the ``2-jets 0-btags'' category is highly enriched in W+light jet events. The modelling of $\ttbar$ background is checked in the ``3-jets 2-btags'' as well as the ``4-jets'' categories.
In general, the event yields are reasonably well reproduced by the simulation within the systematic uncertainties. The shapes of the relevant variables,  $\etalj$ and $\mt$, and the input variables of the NN and BDT analyses show good agreement between data and simulation.

Table ~\ref{tab:yield} shows a difference between the total observed and expected
yields for the $\etalj$ analysis. This difference can be attributed to excesses in data for the Wb+$X$ and
Wc+$X$ processes.
The ATLAS collaboration reported in Ref.~\cite{wbatlas} that the fiducial W+b-jet cross section in the lepton and one or two jets
final state is a factor of 2.1 larger than the NLO prediction, but is still consistent at the level of 1.5 standard deviations with this SM prediction.

Motivated by these observed excesses in comparison to the SM NLO calculations, the $\etalj$ analysis determines the W+jets background yield and
$\etalj$ distribution from data.
The $\etalj$ distribution for W+jets process is extracted from the SB by subtracting the $\etalj$ distribution of all other
processes from the data. The event yield and $\etalj$ distributions used for these subtractions are taken from simulations of $\ttbar$, single-top-quark $s$- and tW- channels, and diboson production. The QCD multijet event yield and $\etalj$ distribution are extracted from data and extrapolated to the SB as described in Section~\ref{sec:qcd}.
The $\etalj$ distribution for W+jets processes in the SB is therefore used in the SR for the signal extraction procedure
(see Section~\ref{sec:etalq}), assuming that the shapes in the SB and SR are compatible with each other.
For the muon channel, the compatibility of the distributions in the two regions has been verified through a Kolmogorov--Smirnov compatibility test, yielding a $p$-value
of 0.47, and a $\chi^2$ test, yielding a $p$-value of 0.63. For the electron channel, the Kolmogorov--Smirnov compatibility test has a $p$-value of 0.51 and the $\chi^2$ test a $p$-value of 0.60.
The stability of the extracted shape has been tested by varying the sample composition in terms of $\ttbar$ and signal fractions by $20\%$ and $100\%$,
respectively. The extracted shapes are compatible with a $p$-value greater than $0.9$ in both cases.

\section{The \texorpdfstring{$\etalj$}{eta[j']} Analysis}
\label{sec:etalq}

\begin{figure}[!ht]
 \begin{center}
	  \includegraphics[width=\cmsFigWidth]{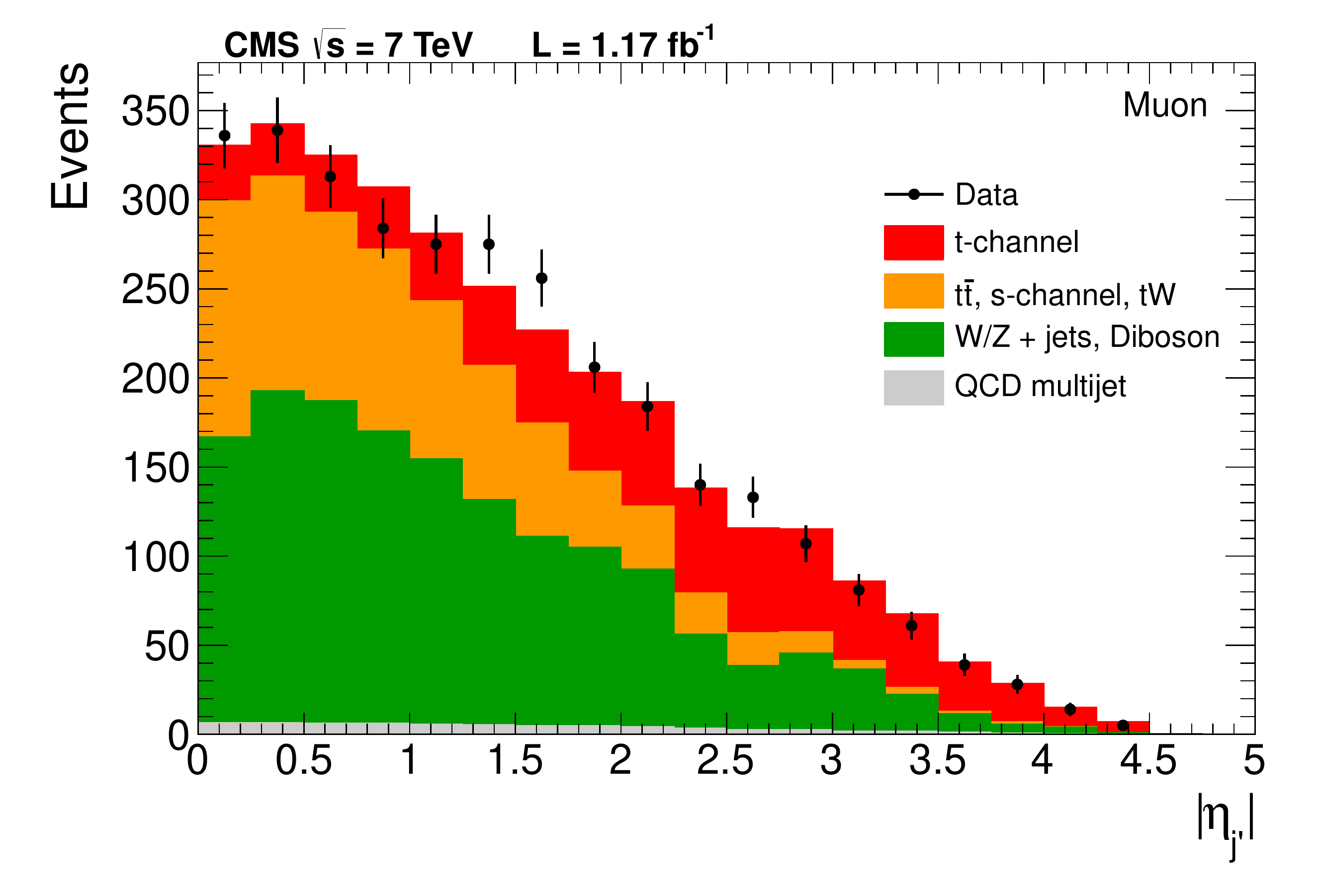}
  \includegraphics[width=\cmsFigWidth]{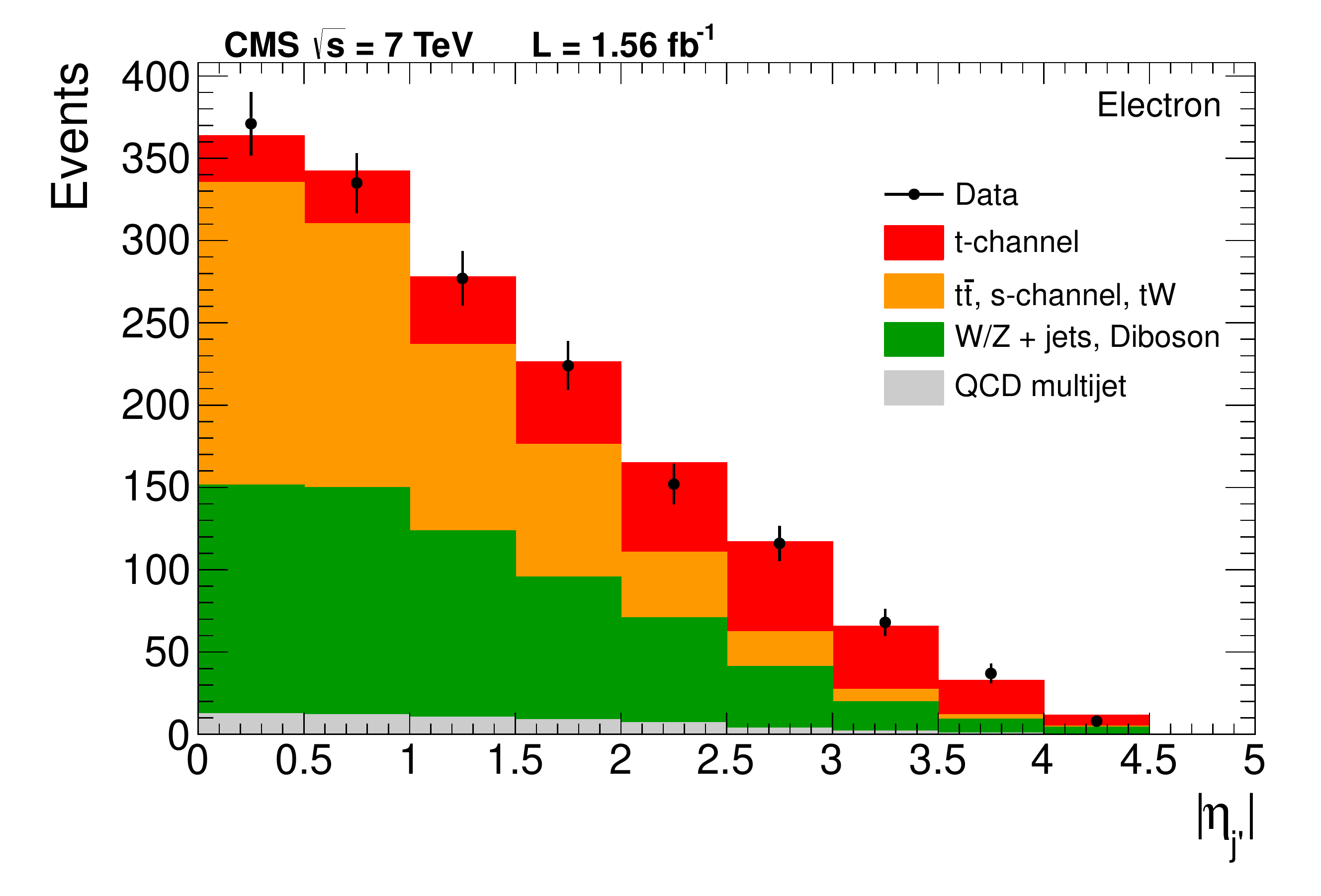}
 \caption{\label{fig:fitresult}{} Result of a simultaneous fit to \etalj in the muon (left) and electron (right) decay channels.}
 \end{center}
\end{figure}

\begin{figure}[!ht]
 \begin{center}

  \includegraphics[width=\cmsFigWidth]{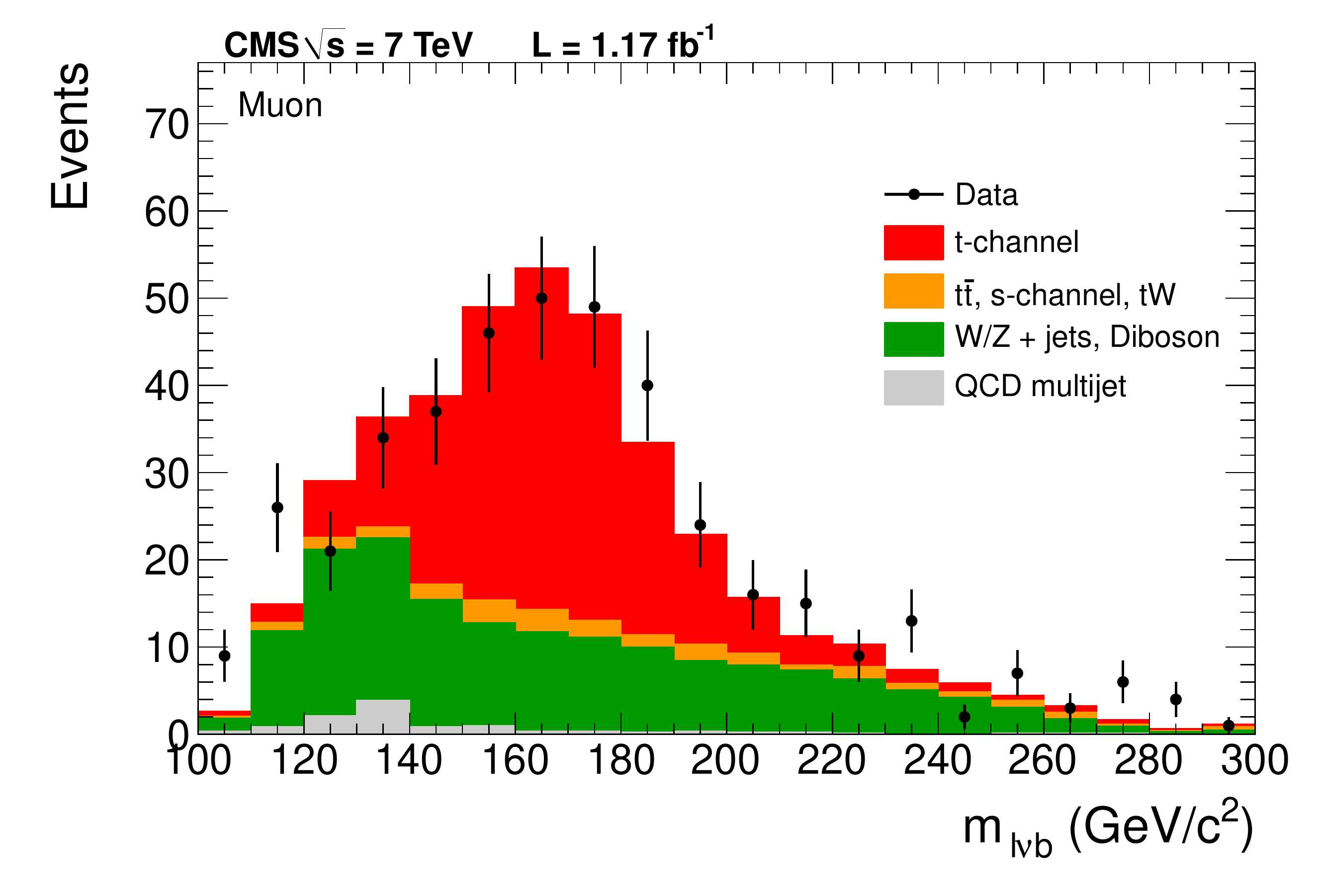}
  \includegraphics[width=\cmsFigWidth]{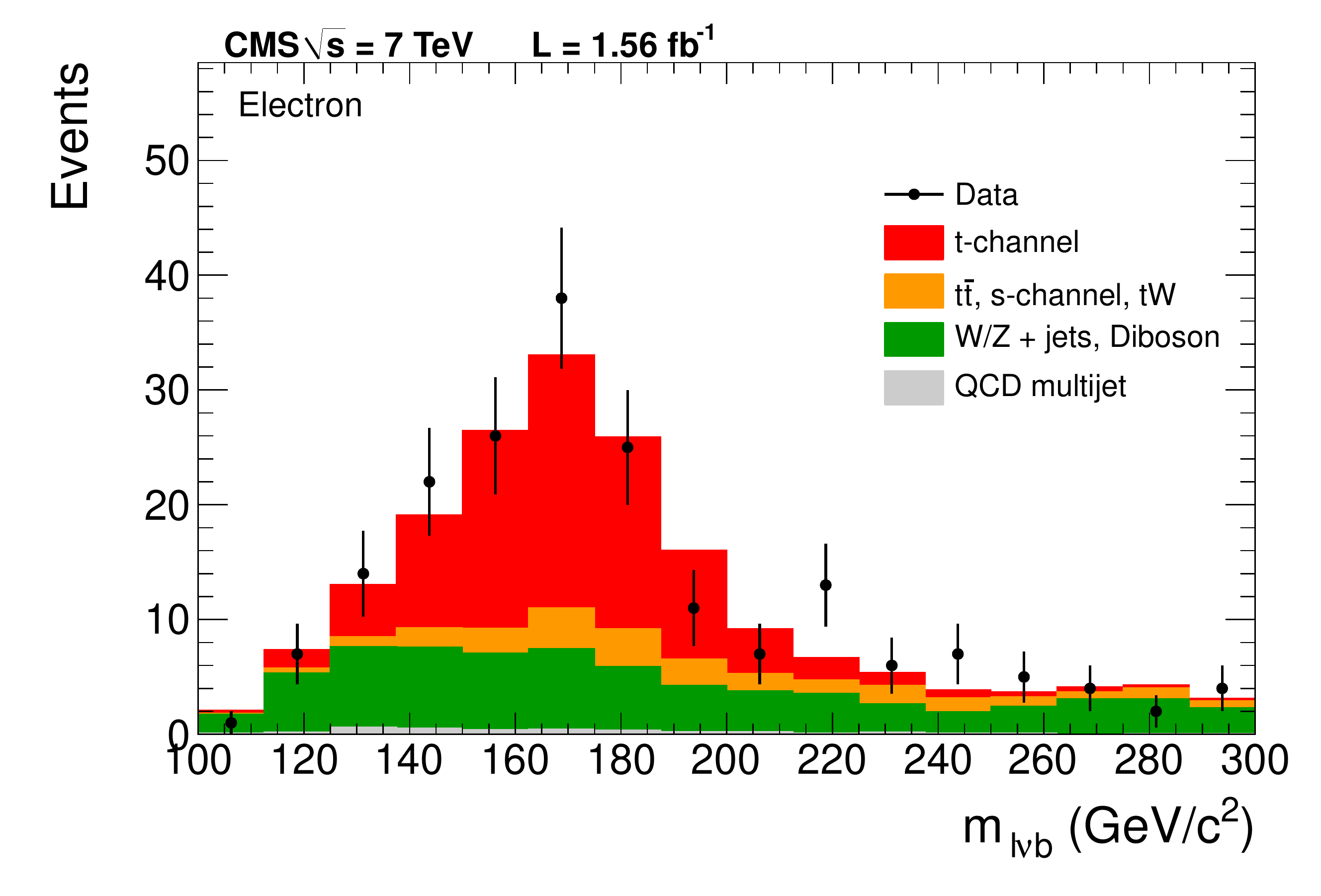}
  \caption{\label{fig:fitstopmass}{} Distributions of $\mt$ requiring $\etalj > 2.8$, for muons (left) and electrons (right), obtained by normalising
each process yield to the value from the fit. Because of limited simulated data, the background distribution is smoothed by using a simple spline curve.}
 \end{center}
\end{figure}

The signal yield is extracted using a maximum-likelihood fit to the observed distribution of $\etalj$.
The signal distribution for the fit is taken from simulation. The W/Z+jets component of the background is normalised to the value obtained from the
extraction procedure described in Section~\ref{sec:whfextraction}, and then added to the diboson processes, resulting in the electroweak
component of the background for the fit. The signal and the electroweak components are unconstrained in the fit, whereas the QCD multijet component is fixed
to the result determined in Section~\ref{sec:qcd}. A Gaussian constraint is applied to $\ttbar$ and other top-quark backgrounds.
Figure~\ref{fig:fitresult} shows the distribution of $\etalj$ obtained from the fit.
Figure~\ref{fig:fitstopmass} shows the distribution of the reconstructed top-quark mass
$\mt$ normalised to the fit results, restricting to the highly enriched region of single-top-quark events for $\etalj > 2.8$.
\begin{figure}[!ht]
 \begin{center}
  \includegraphics[width=\cmsFigWidth]{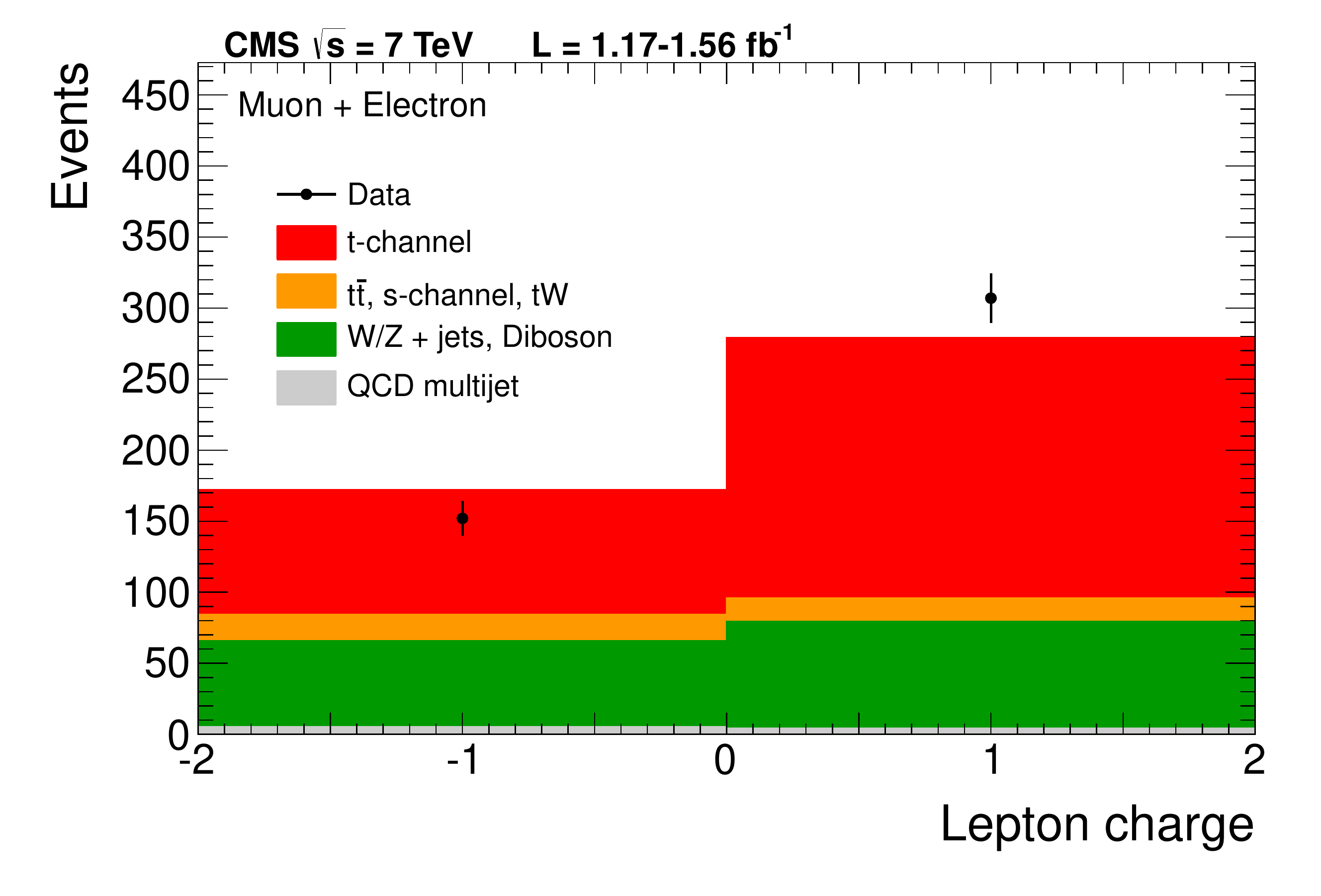}
  \includegraphics[width=\cmsFigWidth]{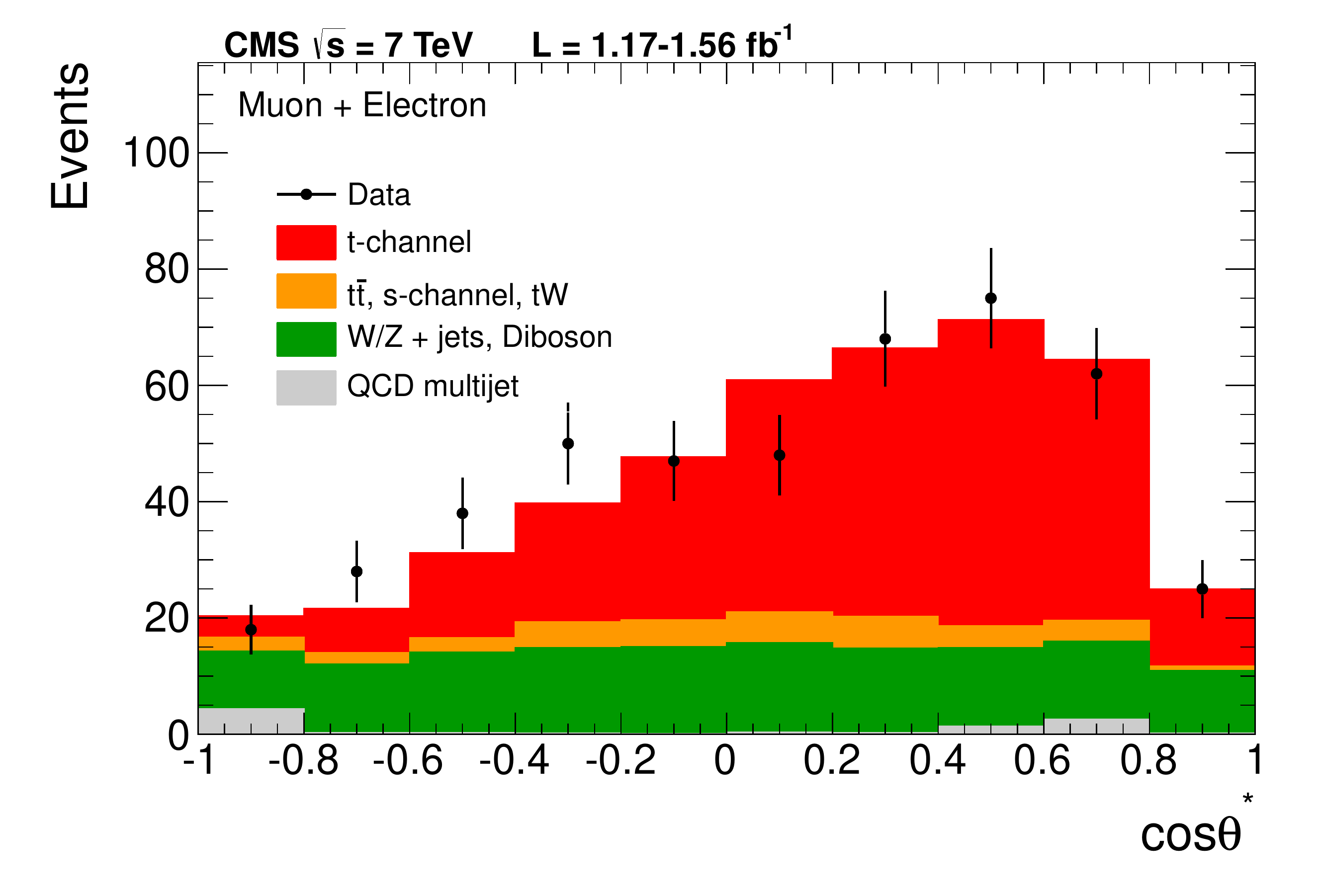}
 \end{center}
  \caption{\label{fig:stop_features}{} Distinct single-top-quark $t$-channel features in the SR for $\etalj > 2.8$, for the electron and muon final states combined. The charge of the lepton (left) and $\cos\theta^*$ (right). All processes are normalised to the fit results. Because of limited simulated data, the background distribution is smoothed by using a simple spline curve (right).}
\end{figure}

Single-top-quark $t$-channel production at the LHC is expected to be characterised by two features. First, the top-quark cross section is about a factor two larger than the top-anti-quark cross section \cite{Kidonakis:2011wy}. This can be experimentally accessed via the charge of the muon or electron. Second, top quarks are almost 100\% polarised with respect to a certain spin axis due to the V--A nature of the couplings. This can be studied via the $\cos\theta^*$ distribution \cite{Motylinski:2009kt}, where $\theta^*$ is defined as the angle between the charged lepton and the non-b-tagged jet, in the reconstructed top-quark rest frame. The observed charge asymmetry and the $\cos\theta^*$ distribution are presented in Fig.~\ref{fig:stop_features} for muon plus electron events in the SR, for $\etalj > 2.8$.

\section{Neural Network Analysis}
\label{sec:nn}

In the NN analysis, several kinematic variables, which are characteristic of SM
single-top-quark production, are combined into a single discriminant by applying an NN technique.
The {\sc NeuroBayes} package~\cite{Feindt:2004, Feindt:2006pm} used for this NN analysis
combines a three-layer feed-forward NN with a complex, but robust, preprocessing.
To reduce the influence of long tails in distributions, input variables are
transformed to be Gaussian distributed.
In addition, a diagonalisation and rotation transformation is performed such
that the covariance matrix of the transformed variables becomes a unit matrix.
To obtain good performance and to avoid overtraining, the NN uses Bayesian
regularisation techniques for the training process.
The network input layer consists of one input node for each input variable plus one bias node. The hidden layer is adapted to this particular analysis and consists of one more node than the input layer. The output node gives a continuous discriminator output in the interval [$-1$, $1$].
For the training of the NN, after applying the full event selection as described in
Section~\ref{sec:selection}, simulated samples of signal $t$-channel single-top-quark events and background events from $\ttbar$, W+jets, and Z+jets samples are used.
The ratio of signal to background events in the training is chosen to be 50:50 and
the background processes are weighted according to the SM prediction, as outlined in
Section~\ref{sec:bkg}.
The NN is trained such that $t$-channel single-top-quark events tend to have
discriminator values close to 1, while background events tend to have discriminator
values near $-1$. Because of their different event selections,
separate neural networks are trained for muon and electron events.

During the preprocessing, the training variables are ranked by the significance
of their correlation to the target discriminator output.
The correlation matrix of all training variables and the target value is calculated. The variable with the smallest
correlation to the target is removed and the loss of correlation is calculated.
This is repeated until the correlation of all variables with the target is determined.
The significance of a variable is calculated by dividing the loss of correlation with the
target by the square root of the sample size. In order to select variables which
contain information that is not already incorporated by other variables, a
selection criterion of $\ge 3\,\sigma$ on the significance has been chosen. A set of 37 variables remains for the muon channel when applying this selection criterion. For the electron channel, a set of 38 variables remains.
The validity of the description of these input variables and the output of the NN
discriminant is confirmed in data with negligible signal contribution.
Furthermore, it is verified with a bootstrapping technique~\cite{efron93bootstrap}
that the bias of the cross section measurement due to a possible overtraining of the
NN is negligible.

\begin{figure}[htbp]
\includegraphics[width=\cmsFigWidth]{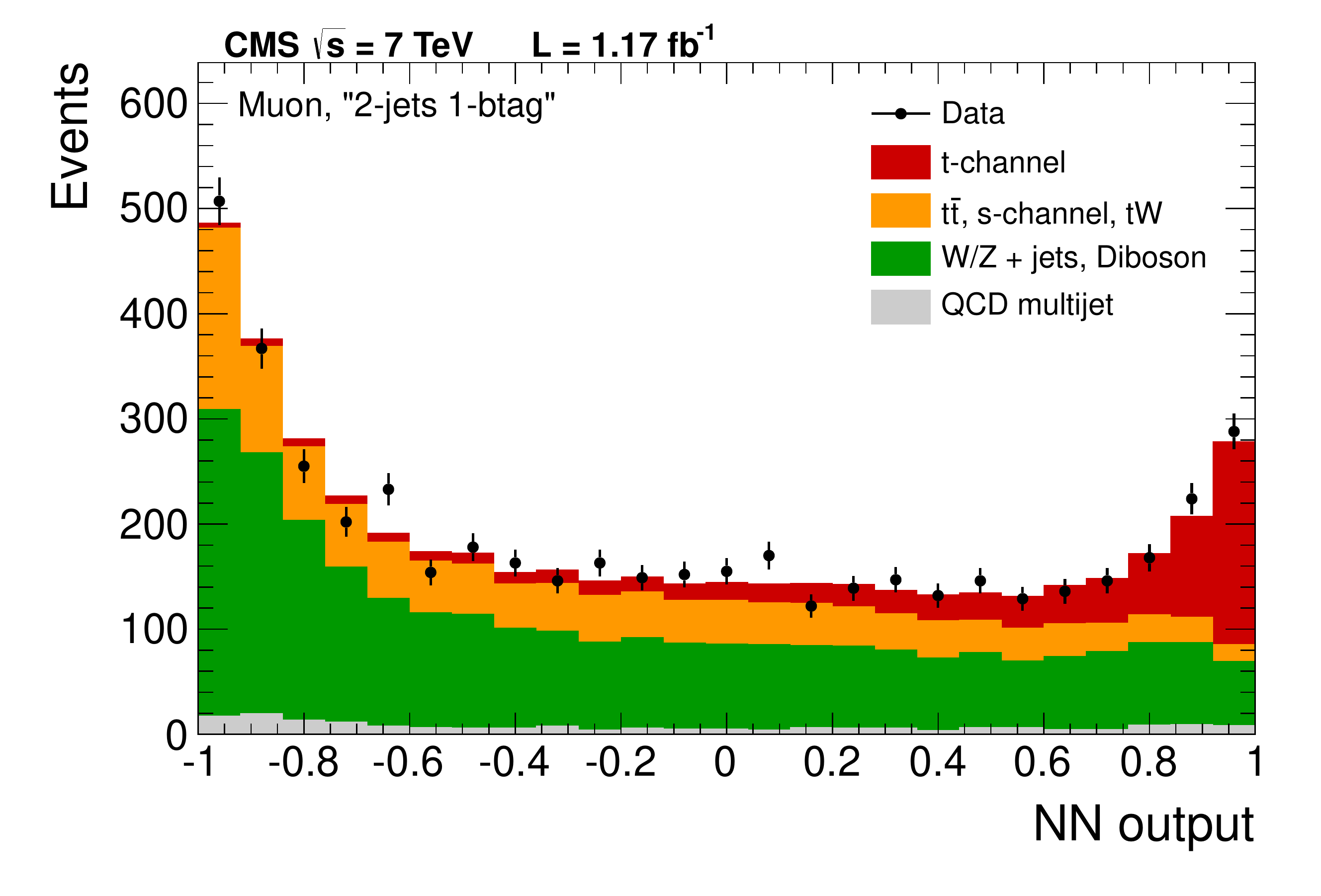}
\includegraphics[width=\cmsFigWidth]{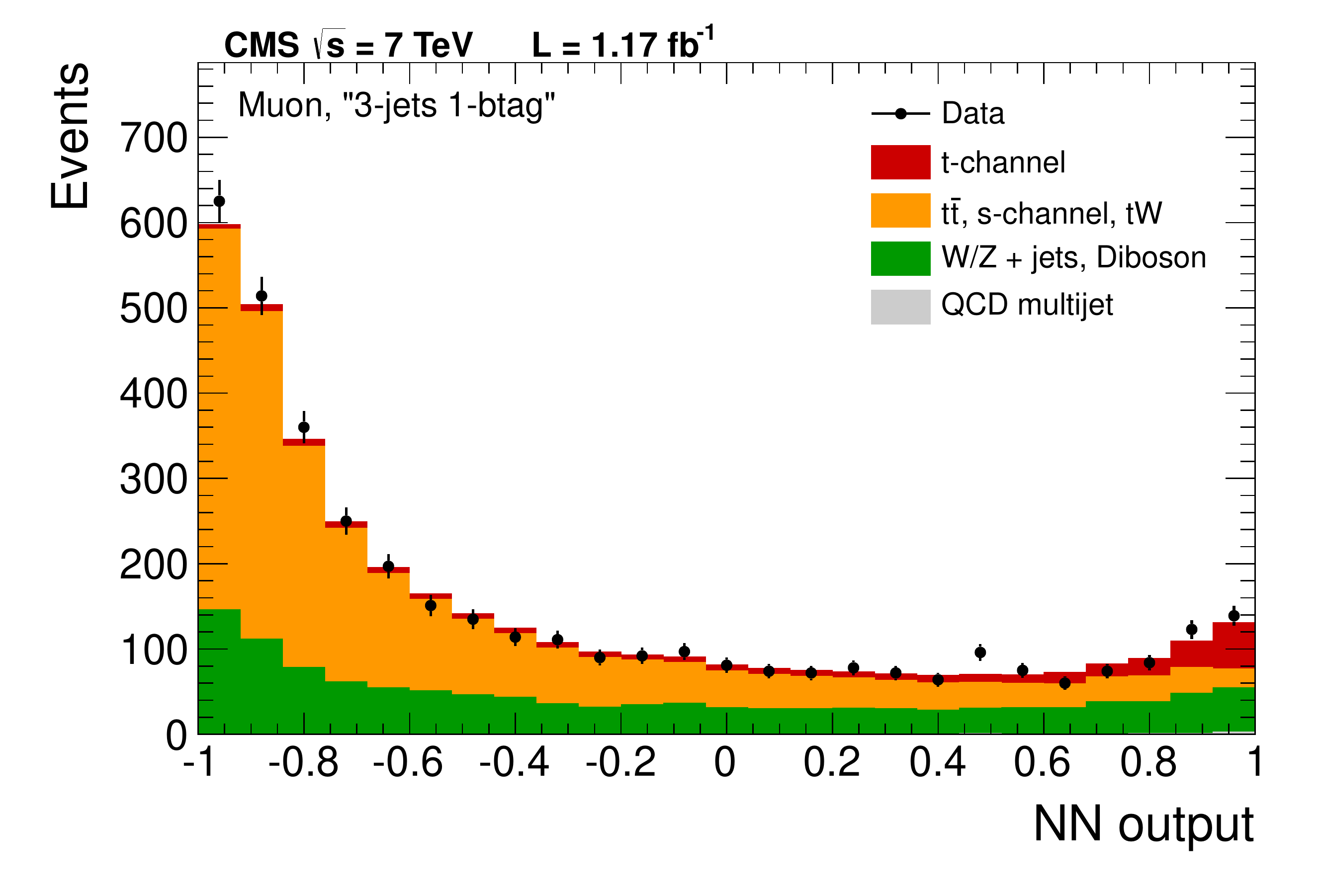}
\caption{Distributions of the NN discriminator output in the muon channel for the ``2-jets 1-btag''  (left) and ``3-jets 1-btag''  (right) categories. Simulated signal and background contributions are scaled to the best fit results.}
\label{Fig:nnstack_mu}
\end{figure}

\begin{figure}[htbp]
\includegraphics[width=\cmsFigWidth]{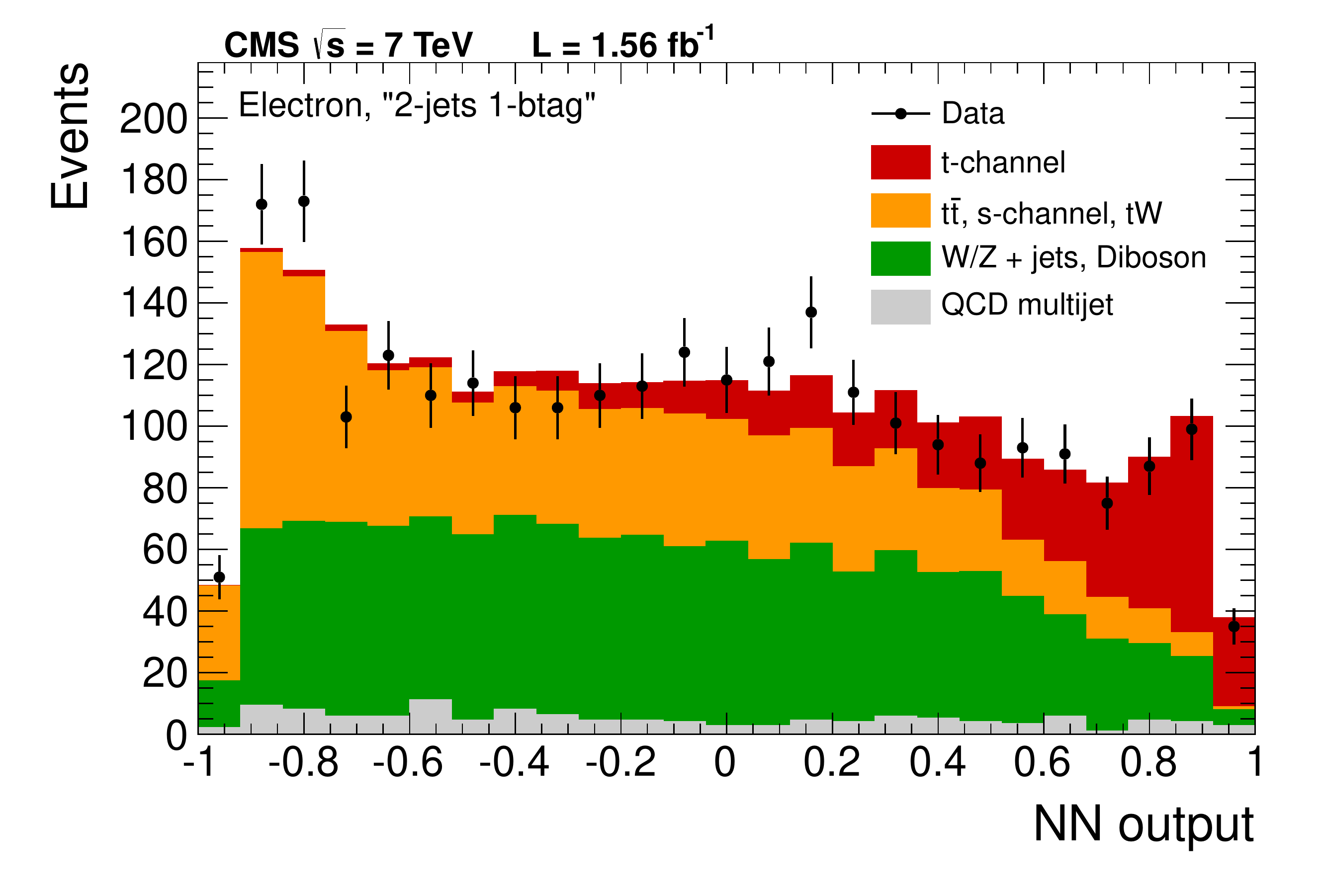}
\includegraphics[width=\cmsFigWidth]{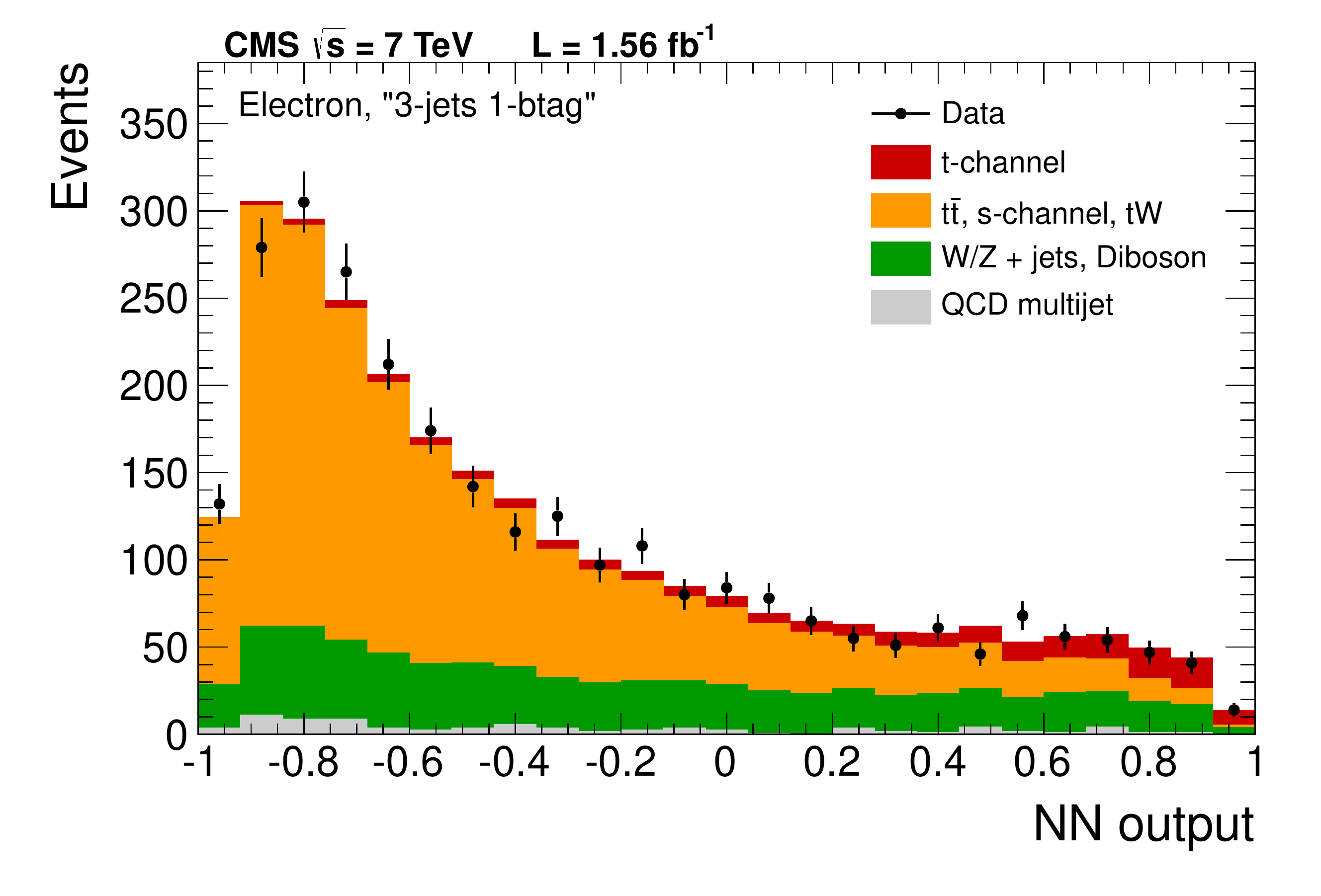}
\caption{Distributions of the NN discriminator output in the electron channel for the ``2-jets 1-btag''  (left) and ``3-jets 1-btag''  (right) categories. Simulated signal and background contributions are scaled to the best fit results.}
\label{Fig:nnstack_ele}
\end{figure}

\begin{figure}[htbp]
\includegraphics[width=\cmsFigWidth]{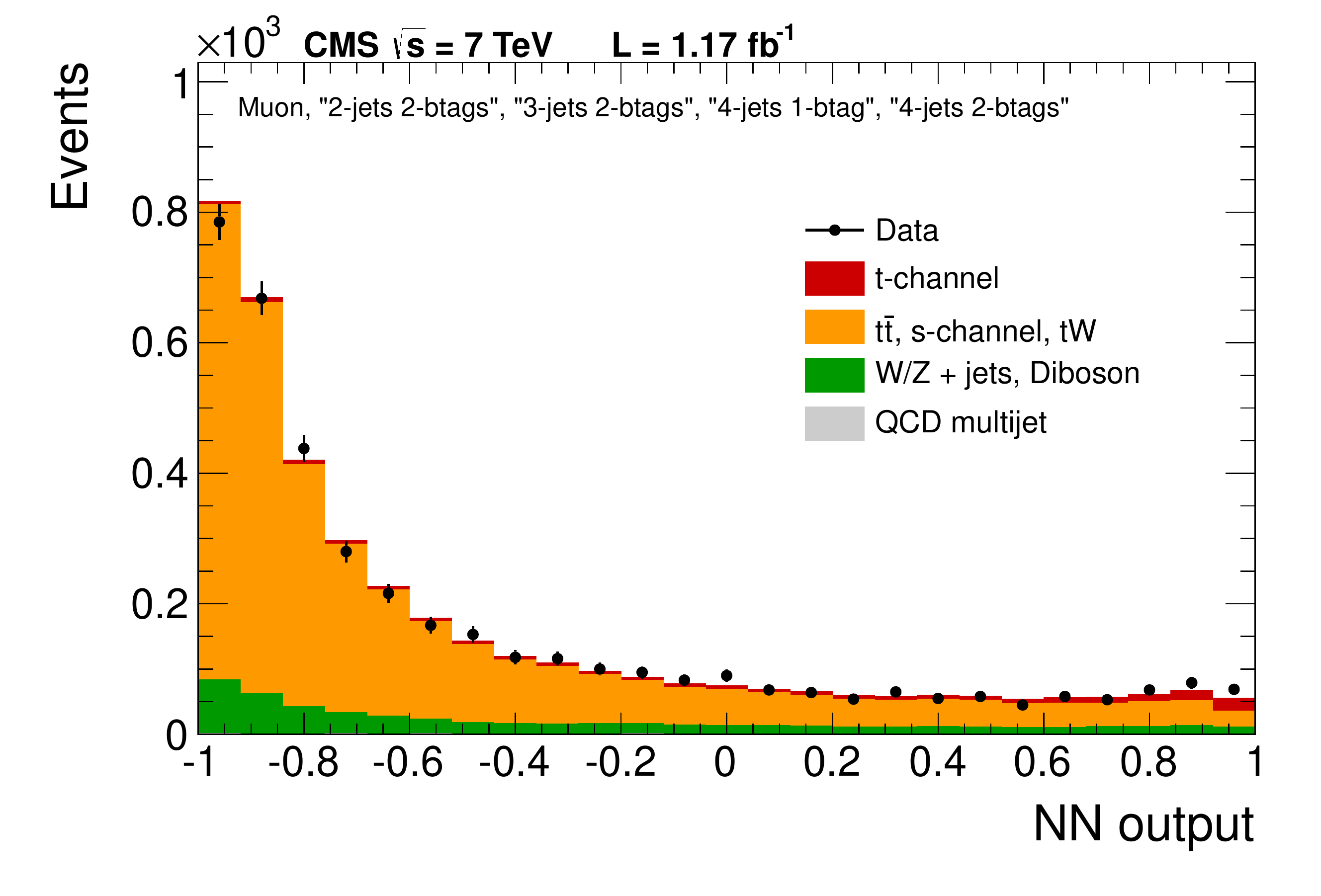}
\includegraphics[width=\cmsFigWidth]{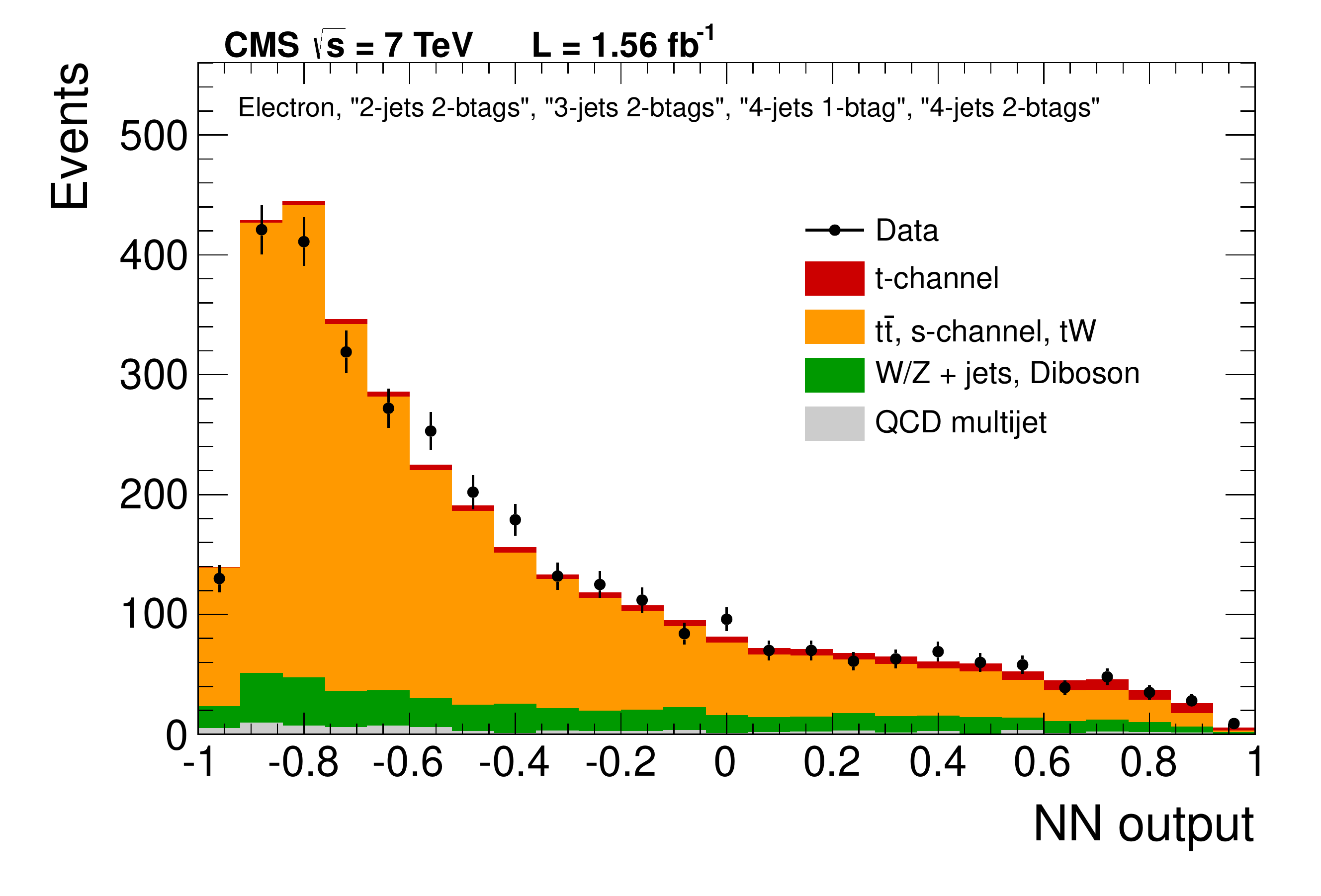}
\caption{Distributions of the NN discriminator output in the background dominated region. All events from the signal depleted categories ``2-jets 2-btags'', ``3-jets 2-btags'', ``4-jets 1-btag'', and ``4-jets 2-btags'' are combined for the muon channel (left) and the electron channel  (right). Simulated signal and background contributions are scaled to the best fit results.}
\label{Fig:nnstack_bg}
\end{figure}
The variables with the highest ranking in both networks are $\etalj$, $\mT$, the invariant mass of the two leading jets, and the total transverse energy of the event.

The distributions of the NN discriminator in the ``2-jets 1-btag'' and ``3-jets 1-btag'' signal categories are shown in Fig.~\ref{Fig:nnstack_mu} for the muon channel and in
Fig.~\ref{Fig:nnstack_ele} for the electron channel. For the remaining categories ``2-jets 2-btags'', ``3-jets 2-btags'', ``4-jets 1-btag'', and ``4-jets 2-btags'', the events are combined into one plot in Fig.~\ref{Fig:nnstack_bg} for the muon channel on the left and for the electron channel on the right.
\section{Boosted Decision Trees Method}
\label{sec:bdt}

The BDT method was previously used in the first CMS measurement of
the $t$-channel cross section \cite{Chatrchyan:2011vp}.
The current analysis uses a significantly larger data sample. It has been optimised using a ``blind'' analysis with the optimisation based exclusively on regions where the signal contribution is negligible and all the selections frozen before the signal region has been looked at. It further provides an increase of the measurement sensitivity and a reduction of systematic uncertainties. The BDT analysis was designed following Ref.~\cite{BDT:Vispa}.

The adopted BDT algorithm constructs 400 decision trees using the Adaptive Boosting algorithm as implemented in Ref.~\cite{tmva}.
The BDT training is carried out separately for the electron and muon final states, for each of the single-top-quark ``2-jets 1-btag'' and ``3-jets 1-btag'' signal-enriched regions, to provide a total of four BDTs. The background processes are input to the training according to their theoretical cross
section predictions as outlined in Section~\ref{sec:bkg}.
One third of the simulated signal and background samples are used for the training,
one third are used to verify the performance of the trained BDT, while the remaining simulated data
provide an unbiased sample used for the cross section evaluation.

Eleven observables reconstructed in the detector are chosen based on their power to discriminate between signal and background events.

The adopted variables are the lepton transverse momentum;
the pseudorapidities of the most forward non-b-tagged jet and of the jet with the highest transverse momentum;
the invariant mass of all reconstructed jets in the event; the angular separation
between the two leading jets;
the sums of the hadronic energy and of the hadronic transverse energy;
the reconstructed top-quark mass using the jet with the highest b-tag discriminator;
the reconstructed top-quark mass using the jet giving the reconstructed top-quark mass closest to
172\GeVcc; the cosine of the angle between the reconstructed
W boson, in the rest frame of the sum of four-vectors of the W boson and leading jet, and the sum of four-vectors of the
W boson and leading jet; and the sphericity of the event.
The validity of the description of these input variables and the output of the BDT classifiers is confirmed in data with negligible signal contribution using a Kolmogorov--Smirnov test.

The QCD multijet background evaluation is described in Section~\ref{sec:bkg}.
The determination of the single-top-quark production cross section in the $t$-channel, including the treatment of statistical and systematic
uncertainties, is performed by using the classifier distributions in all 12 analysis categories simultaneously.

\begin{figure}[htbp]
\begin{center}
\includegraphics[width=\cmsFigWidth]{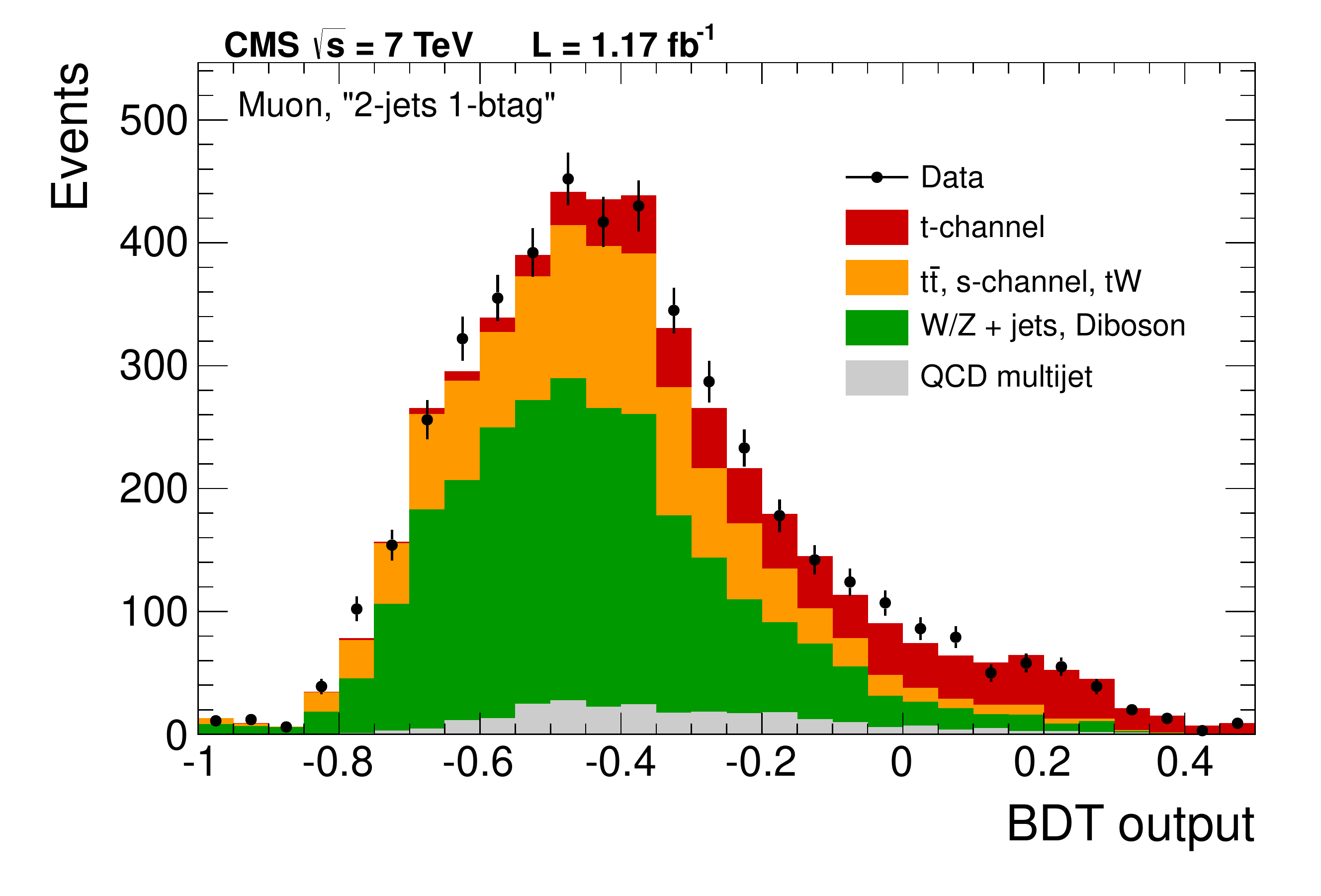}
\includegraphics[width=\cmsFigWidth]{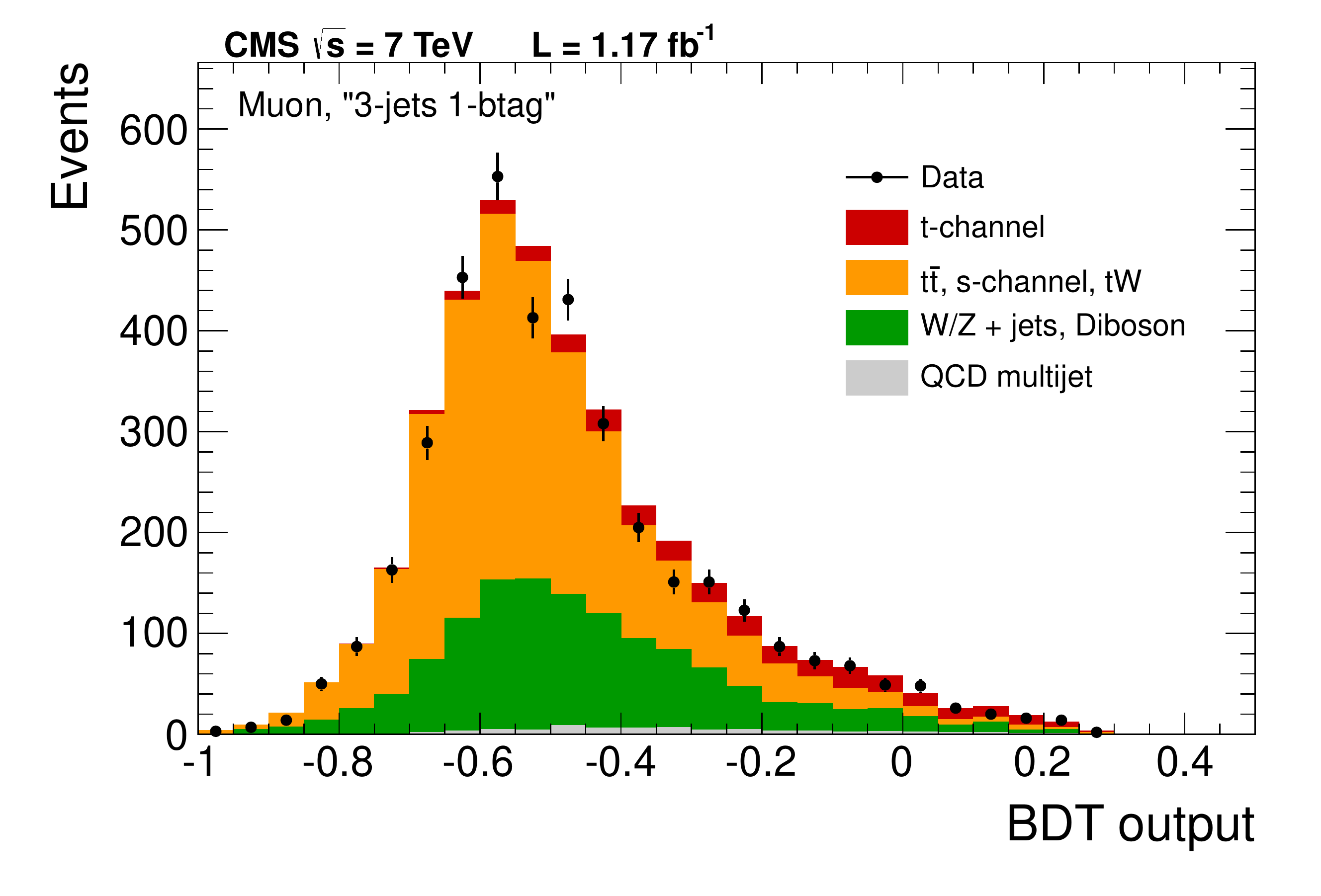}
\caption{Distributions of the BDT discriminator output in the muon channel for the ``2-jets 1-btag''  (left) and ``3-jets 1-btag''  (right) categories.  Simulated signal and background contributions are scaled to the best fit results.}
\label{fig:bdt_mu}
\end{center}
\end{figure}

\begin{figure}[htbp]
\begin{center}
\includegraphics[width=\cmsFigWidth]{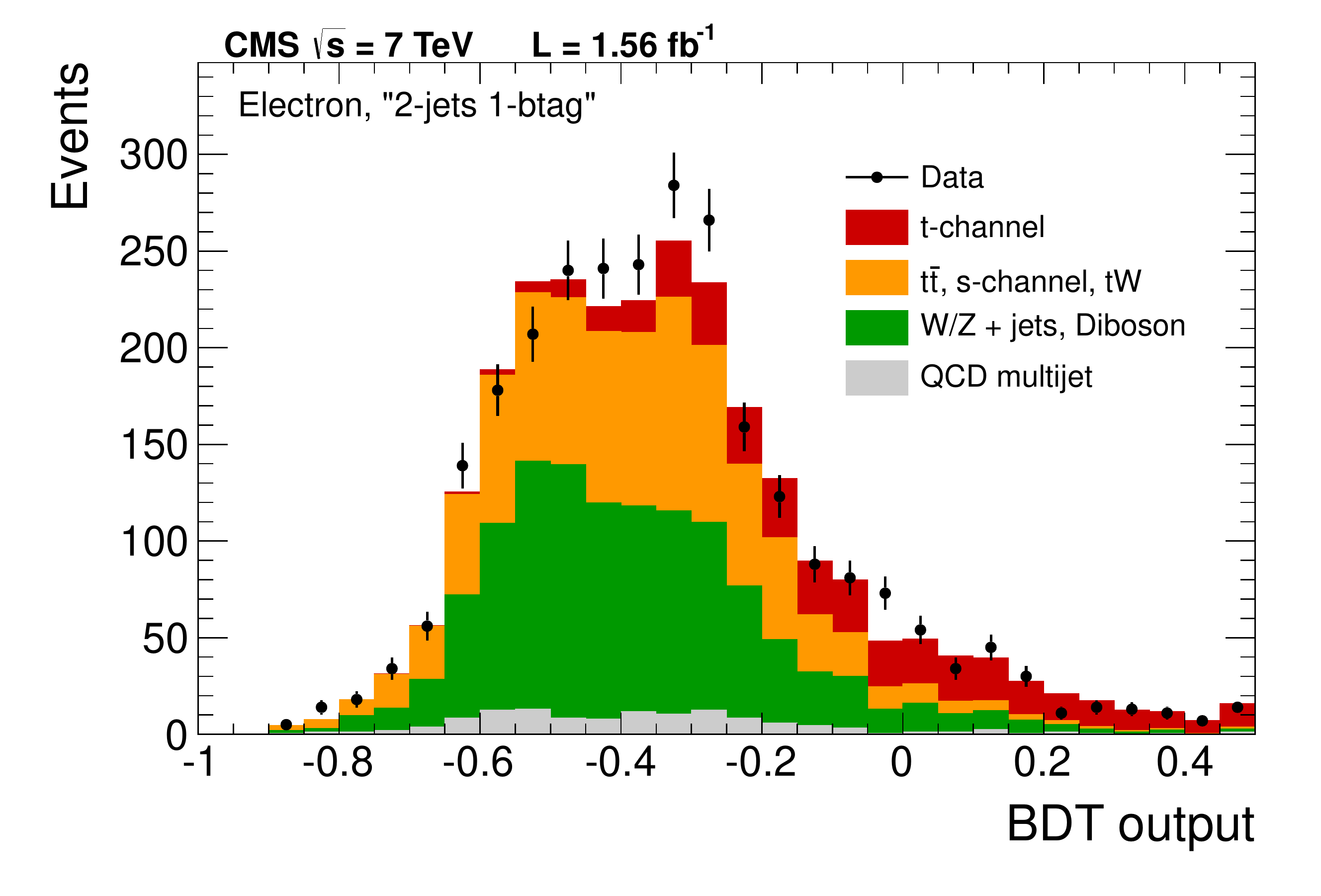}
\includegraphics[width=\cmsFigWidth]{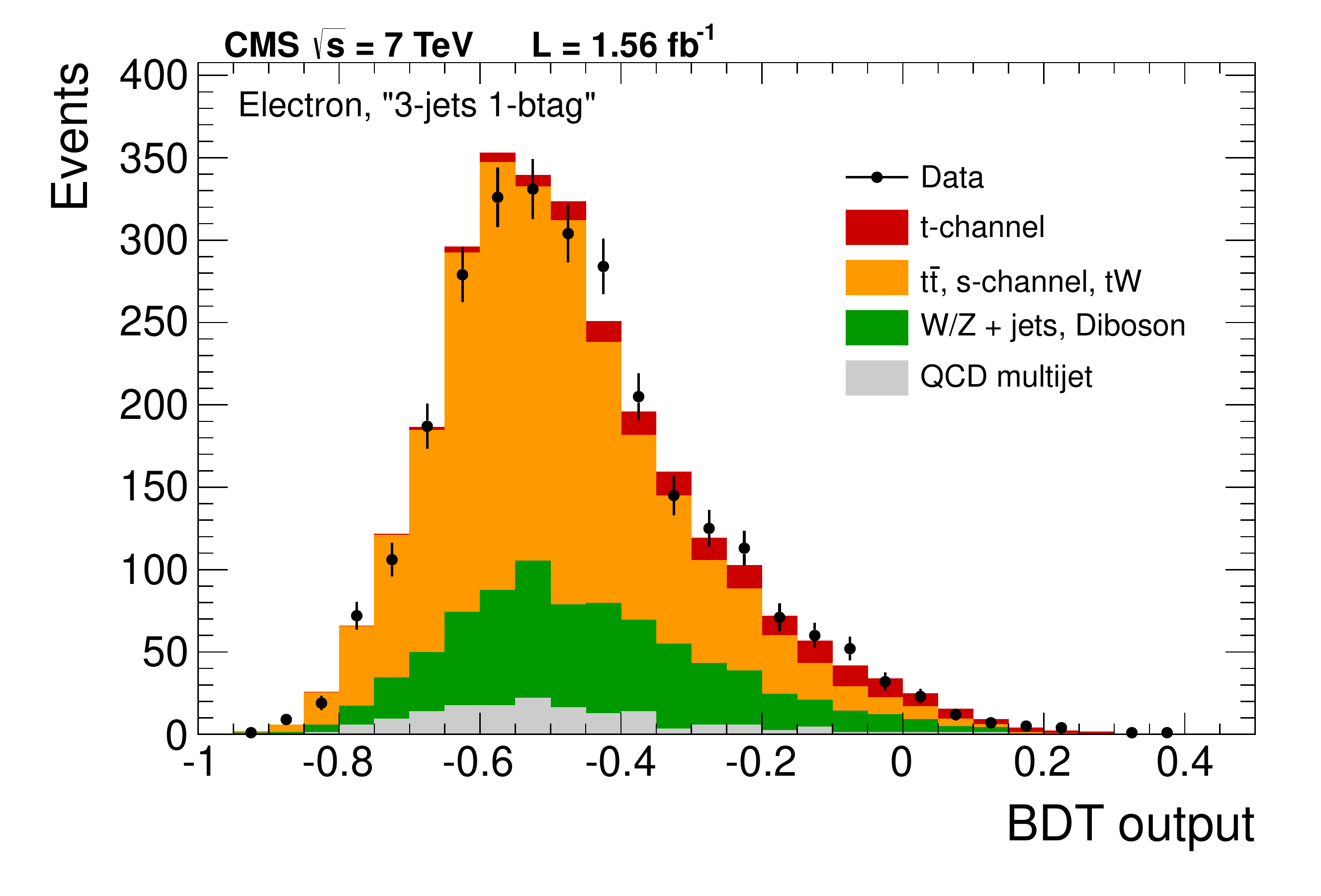}
\caption{Distributions of the BDT discriminator output in the electron channel for the ``2-jets 1-btag''  (left) and ``3-jets 1-btag''  (right) categories. Simulated signal and background contributions are scaled to the best fit results.}
\label{fig:bdt_e}
\end{center}
\end{figure}

\begin{figure}[htbp]
\begin{center}
\includegraphics[width=\cmsFigWidth]{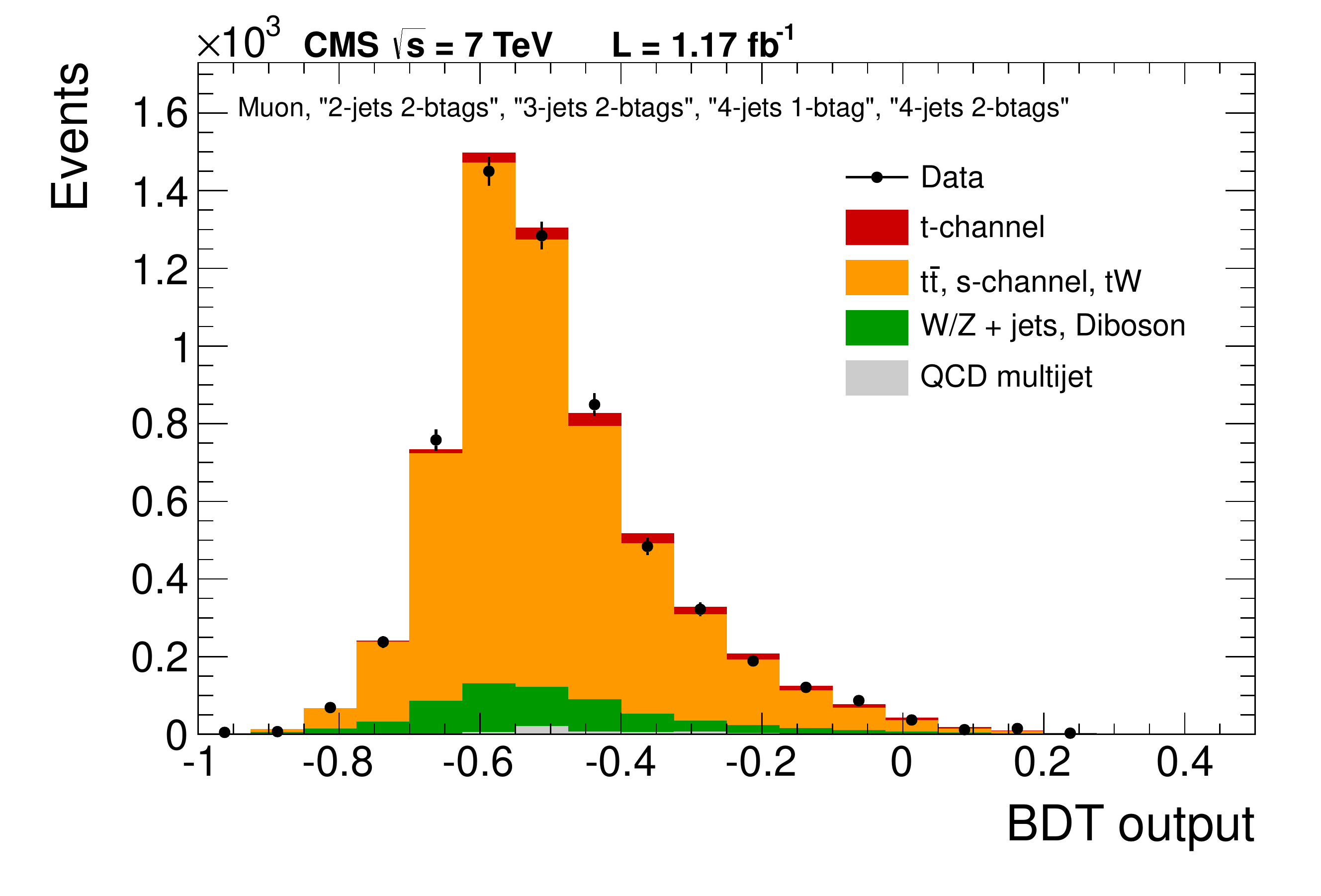}
\includegraphics[width=\cmsFigWidth]{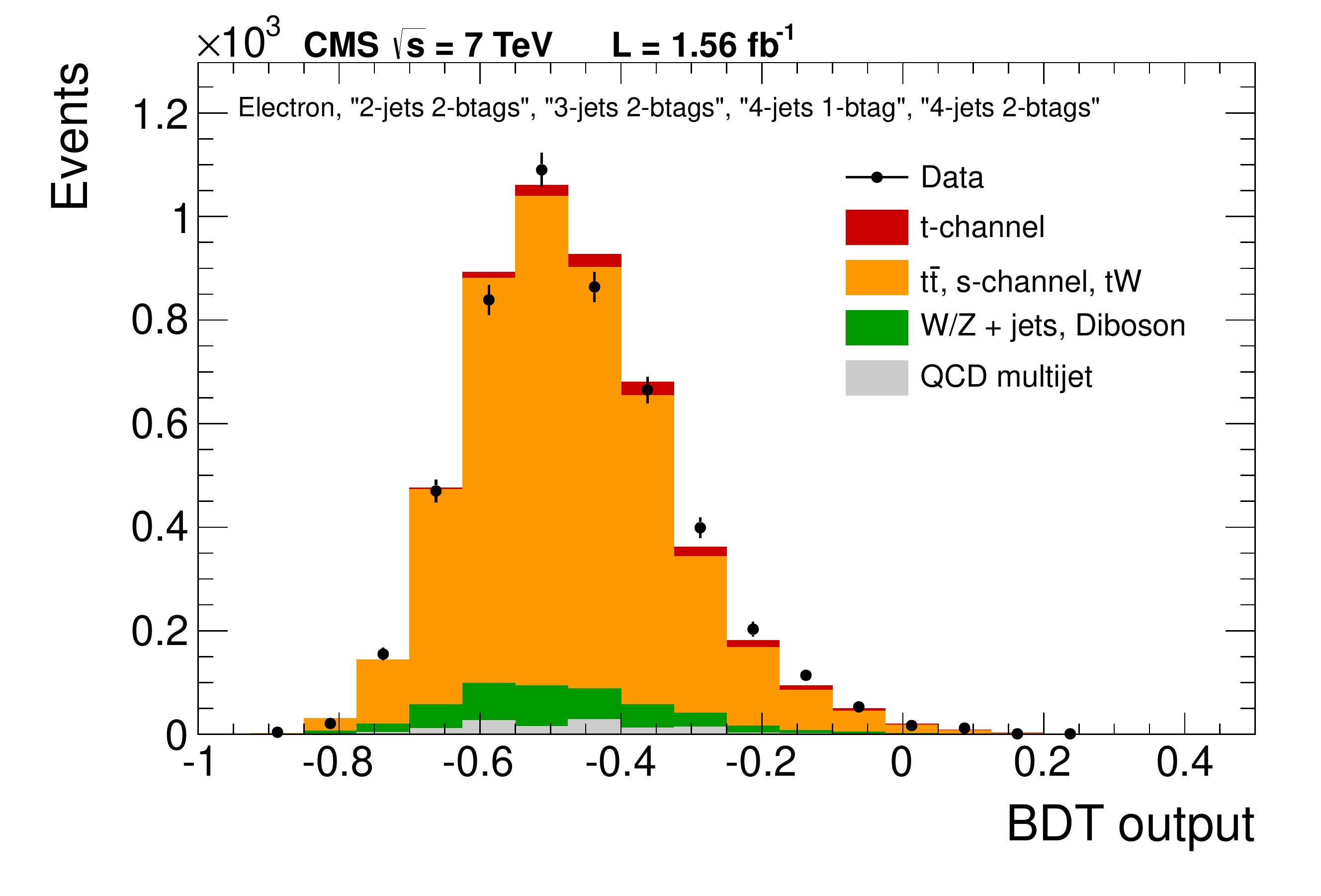}
\caption{Distributions of the BDT discriminator output in the signal depleted categories for the muon channel (left) and electron channel (right).  Simulated signal and background contributions are scaled to the best fit results.}
\label{fig:bdt_controlregions}
\end{center}
\end{figure}

The measured distributions of the classifier outputs in the "2-jets 1-btag" and "3-jets 1-btag" signal enriched categories are shown separately for the muon and electron channel in Figs.~\ref{fig:bdt_mu} and~\ref{fig:bdt_e}.
The measured distributions of the classifier outputs in the "2-jets 2-btags","3-jets 2-btags", "4-jets 1-btag", and "4-jets 2-btags" signal depleted categories are shown separately for the muon and electron channels in Fig.~\ref{fig:bdt_controlregions}.

\section{Determination of the Cross Section with Multivariate Analyses}
\label{sec:bayes}
The NN and BDT analyses employ a Bayesian approach~\cite{Jaynes2003} to measure the single-top-quark production cross section.
The signal cross section is determined simultaneously from the data distributions of the corresponding multivariate discriminator, modelled as $1$-dimensional histograms, in the different categories.
The distributions for signal and backgrounds are taken from simulation, except for the QCD multijet distribution, which is derived from data.
The signal yield is measured in terms of the signal strength $\mu$ which is defined as the actual cross section divided by the SM prediction.
The probability to observe a certain dataset $p(\mathrm{data}\vert\mu)$ is related
to the posterior distribution $p(\mu\vert\text{data})$ of $\mu$ by using Bayes' theorem
\begin{align}
p(\mu\vert\text{data}) \propto {p(\text{data}\vert\mu)\cdot\pi(\mu)}\,\nonumber
\end{align}
where $\pi(\mu)$ denotes a uniform prior distribution of the signal strength.

Experimental uncertainties and cross-section uncertainties are included by introducing additional parameters $\vec\theta$ in the statistical model, extending the probability to observe a certain dataset to
$p^\prime(\mathrm{data}\vert\mu,\vec{\theta})$, which depends on the nuisance parameters $\vec\theta$. This approach
allows the nuisance parameter values to be constrained by data, thus reducing the respective uncertainties.

The $m$ systematic uncertainties are included via nuisance parameters
$\vec{\theta}=(\theta_1, \cdots , \theta_m)$, which contain $k$ normalisation parameters
and $m-k$ parameters influencing the shape of the simulated distributions.
For normalisation uncertainties $i=1,\cdots, k$, the priors $\pi_i(\theta_i)$ are log-normal
distributions. The medians are set to the corresponding cross sections and the widths to their uncertainties.
The priors for shape uncertainties are normal distributions. For each shape uncertainty, two shifted simulated discriminator distributions are derived, varying the corresponding uncertainty by $\pm 1\sigma$. Here, the parameter $\theta_i$ is used to interpolate between the nominal and shifted histograms.

The posterior distribution for the signal strength $\mu$, $p(\mu\vert\mathrm{data})$, is obtained by integrating the
$(m+1)$-dimensional posterior in all nuisance parameters $\vec{\theta}$:
\begin{align}
p(\mu\vert\mathrm{data}) \propto \int p^\prime(\mathrm{data}\vert\mu,\vec{\theta})\cdot \pi(\mu)\pi(\vec{\theta})\; \mathrm{d}\vec{\theta}\,.\nonumber
\end{align}
This integration, also called \emph{marginalisation}, is performed using the Markov chain MC method as implemented in the package \textsc{Theta}~\cite{theta}.
The central value is extracted from this distribution and used as $\mu$. The central $68\%$ quantile is taken as the total marginalised uncertainty in the measurement. It includes the statistical uncertainty and the $m$ systematic uncertainties.

An estimate of the statistical uncertainty is taken from the total marginalised uncertainty after subtracting, in quadrature, the individual contributions of marginalised uncertainties. To obtain the individual contribution of each systematic uncertainty, the signal extraction procedure is repeated with signal and background contributions changed according to the systematic uncertainty. The mean shift of the cross section estimate, with respect to the value obtained in the nominal scenario, is taken as the corresponding impact on the signal cross section measurement.

However, modelling the systematic uncertainties as nuisance parameters requires an assumption for the dependence of $p^\prime$ on $\theta_i$.
This dependence is often unknown for theoretical uncertainties, in particular if they have an effect on the shape of the discriminator distribution. Therefore,
such uncertainties are not included via additional nuisance parameters. Instead, their effect on the cross section measurement is estimated by performing pseudo-experiments as explained above, and their impact is added in quadrature to the total marginalised uncertainty.

In conclusion, systematic uncertainties are included with two different methods. Experimental uncertainties and cross-section uncertainties are included as additional parameters in the statistical model and marginalised. Theoretical uncertainties are not included as additional parameters in the statistical model, but their impact is added in quadrature to the total marginalised uncertainty.

\section{Systematic Uncertainties and Measurement Sensitivity}
\label{sec:syst}

For the $\etalj$ analysis, each systematic uncertainty is evaluated by generating pseudo-experiments, which take into account the effect of the corresponding systematic source on the distribution of $\etalj$ and on the event yield of the physics processes. Pseudo-experiments are generated separately with templates varied by ${\pm}1\,\sigma$ of the corresponding uncertainty. A fit to $\etalj$ is then performed on each pseudo-experiment. The mean shift of the fit results, with respect to the value obtained in the nominal scenario, is taken as the corresponding uncertainty.

In the BDT and NN analyses the experimental systematic uncertainties (excluding the luminosity) are marginalised with the Bayesian method. Theoretical uncertainties, however, are not marginalised, but are estimated by generating pseudo-experiments using separate templates, varied by ${\pm}1\,\sigma$ for the corresponding uncertainty, for each source of systematic uncertainties, and repeating the signal extraction procedure.

A particular uncertainty, which is only present for the $\etalj$ analysis, concerns the extraction of W+jets from data. It is evaluated by generating pseudo-experiments in the SB and repeating the signal extraction procedure and fit to $\etalj$. This method exploits the ansatz that the distribution of $\etalj$ is the same in both the SR and SB. The uncertainty is taken as the root mean square of the distribution of fit results obtained in this way. This uncertainty depends
on the amount of available data in the SB and is uncorrelated between the muons and electron samples.
In addition, alternative $\etalj$ shapes are derived in the simulation by varying the Wb+X and Wc+X fractions of the background
by  factors of  ${\pm}30\%$ independently in the SR and SB regions. The fit procedure is repeated using the new shapes and the maximum difference in the result with respect to the central value is added in quadrature to the other uncertainties.

The following sources of systematic uncertainties are considered in all three analyses.
Differences in the $\etalj$ and the two multivariate analyses are remarked upon where relevant:
\begin{itemize}

\item \textbf{Jet energy scale (JES)}. All reconstructed jet four-momenta in simulated events
are simultaneously varied according to the $\eta$ and $\pt$-dependent
uncertainties on the jet energy scale~\cite{CMS-PAS-JME-10-010}. This variation in jet four-momenta is also propagated to $\MET$.
For multivariate analyses, the complete parameterisation of JES as in~\cite{CMS-PAS-JME-10-010} is considered and all parameters are considered as nuisance parameters and included in the marginalisation procedure described in Section~\ref{sec:bayes}.

\item \textbf{Jet energy resolution}. A smearing is applied to account for the known difference in jet energy resolution with respect to data~\cite{CMS-PAS-JME-10-014}, increasing or decreasing the extra resolution contribution by the uncertainty on the resolution.

\item  \textbf{b tagging}. Both b tagging and misidentification efficiencies in the data are estimated from control samples~\cite{btag}.
Scale factors are applied to simulated samples to reproduce the measured efficiencies. The corresponding uncertainties are propagated as systematic uncertainties. For multivariate analyses, b tagging average scale factors are constrained using the marginalisation procedure described in Section~\ref{sec:bayes}.
The effect of any remaining, unconstrained, b tagging modelling is determined by modelling possible
variations of the b tagging scale factors as a function of $\pt$ and $|\eta|$, using different degrees (from one to five) of Chebyshev polynomials.
The largest observed variation with respect to the nominal result is taken as the systematic uncertainty,
and is found to be negligible.

\item \textbf{Trigger}. Single lepton trigger efficiencies are estimated with a ``tag and probe'' method~\cite{Khachatryan:2010xn} from Drell--Yan data. The efficiencies of  triggers requiring a lepton plus a b-tagged jet are parameterised as a function of the jet $\pt$ and the value of the TCHP b-tag discriminator. The selection efficiencies have been validated using a reference trigger. The uncertainties of the parameterisation and an additional flavour dependency is propagated to the final result.

\item \textbf{Pileup}. The effect of multiple interactions (pileup) is evaluated by reweighting simulated samples to reproduce the expected number of pileup interactions in data, properly taking into account in-time and out-of-time pileup contributions. The uncertainty on the expected number of pileup interactions (5\%) is propagated as a systematic uncertainty to this measurement.

\item \textbf{Missing transverse energy}. The $\MET$ modelling uncertainty is propagated to the cross section measurement. The effect on $\MET$ measurement of unclustered energy deposits in the calorimeters is included.

\item \textbf{Luminosity}. The luminosity is known with a relative uncertainty of $\pm 2.2\%$~\cite{lumi}.

\item \textbf{Background normalisation}. The uncertainties on the normalisation of each background
source are listed below. They are propagated as systematic uncertainties in the $\etalj$
analysis only for dibosons and $s$- and tW-channel single-top-quark processes. The remaining backgrounds are estimated from data. The uncertainty on
$\ttbar$ is used as a Gaussian constraint
in the signal-extraction fit. In the multivariate analyses, normalisation uncertainties are accounted for
as a prior probability density function for the Bayesian inference using a log-normal model.

\begin{itemize}
\item $\ttbar$: ${\pm}15\%$, based on the statistical uncertainties in Ref.~\cite{top-10-003}.

\item Dibosons, single-top-quark $s$- and tW-channels: ${\pm}30\%$, ${\pm}15\%$, ${\pm}13\%$, respectively, based on theoretical uncertainties.

\item W/Z+jets: ${\pm}100\%$, ${\pm}50\%$, and ${\pm}30\%$ are taken for W+b/c flavour jets, W+light flavour jets, and Z+jets, respectively, consistent with previous estimates~\cite{top-10-003}. In the multivariate analyses, the various W+jets processes are considered to be uncorrelated, as are the different jet categories, in order to avoid too many model assumptions.

\item QCD multijet: the normalisation and the corresponding uncertainty is determined from data (see Section~\ref{sec:bkg}).
\end{itemize}

\item \textbf{Limited MC data}. The uncertainty due to the limited amount of MC data in the templates used for the statistical inferences is determined by using the Barlow--Beeston method~\cite{barlow_beeston, bb_light}.

\item \textbf{Scale uncertainty}. The uncertainties on the renormalisation and factorisation scales are studied with dedicated single-top-quark and background samples of W+jets, Z+jets, and $\ttbar$ events. They are generated by doubling or halving the renormalisation and factorisation scale with respect to the nominal value equal to the $Q^2$ in the hard-scattering process.

\item \textbf{Extra parton modelling (matching)}. The uncertainty due to extra hard parton radiation is studied by doubling or halving
the threshold for the MLM jet matching scheme~\cite{mlm} for W+jets, Z+jets, and $\ttbar$ from its default.

\item \textbf{Signal generator}. The results obtained by using the nominal
\POWHEG signal samples are compared with the result obtained using
signal samples generated by \linebreak \COMPHEP. In general, the largest
model deviations occur in the kinematic distributions of the spectator
b quark~\cite{Campbell:2009ss}.
The differences in the transverse momentum distribution of the spectator
b quark, at the generator level, between 4-flavour and 5-flavour scheme
(FS)~\POWHEG~\cite{Frederix:2012dh} are more than a factor two
smaller than the differences between \textsc{powheg-5fs} and \COMPHEP over the whole \pt range.
Pseudo-experiments with simulated \COMPHEP events are generated
and the nominal signal extraction procedure with \textsc{powheg-5fs} templates is repeated. Half of the observed shift of the cross section
measurement is taken as the systematic uncertainty due to signal
generator modelling.

\item \textbf{Parton distribution functions}. The uncertainty due to the choice of the parton distribution functions (PDF) is estimated using pseudo-experiments, reweighting the simulated events with each of the 40 eigenvectors of the {CTEQ6}~\cite{PDF:CTEQ6} PDF set and the central set of {CTEQ10}~\cite{PDF:CTEQ10}, and repeating the nominal signal extraction procedure. For reweighting the simulated events, the \textsc{lhapdf}~\cite{PDF:LHAPDF} package is used.
\end{itemize}

The jet energy scale and jet energy resolution are fully correlated across all samples.
The matching and scale uncertainties are fully correlated between W+jets and Z+jets, but are uncorrelated with $\ttbar$.

Table~\ref{tab:syst_combined} summarises the different contributions to the systematic uncertainty on the combined
(muon and electron) cross section measurement in the three analyses.
For the multivariate analyses, experimental uncertainties and background rates are constrained from
the marginalisation procedure (except for the luminosity), as described in Section~\ref{sec:bayes}. The remaining theoretical and luminosity
uncertainties are added separately in quadrature to the total uncertainty.

\begin{table}
\topcaption{Sources of uncertainty on the cross section measurement.}
\label{tab:syst_combined}
\begin{center}
\begin{tabular}{  l  l | l | c  c c  }
 \hline
\multicolumn{2}{l}{ } & Uncertainty source & NN & BDT & $\etalj$ \\
\hline
\hline
 \multirow{16}{*}{\rotatebox{90}{Marginalised (NN, BDT)}} &  \multirow{9}{*}{\rotatebox{90}{Experimental uncert.}}  &Statistical  & $-6.1$/$+5.5\%$ & $-4.7$/$+5.4\%$ & $\pm 8.5\%$  \\
 & & Limited MC data & $-1.7$/$+2.3\%$ & $\pm 3.1\%$ & $\pm 0.9\% $  \\
 \cline{3-6}
& & Jet energy scale & $ -0.3$/$+1.9\%$ & $ \pm 0.6\%$ & $ - 3.9 $/$ + 4.1  \% $  \\
& & Jet energy resolution & $-0.3$/$+0.6\%$ & $ \pm 0.1\%$ & $ - 0.7 $/$ + 1.2 \% $  \\
& & b tagging & $-2.7 $/$ +3.1\%$ & $ \pm 1.6\%$ & $\pm 3.1\%$ \\
& & Muon trigger + reco. & $-2.2$/$+2.3\%$ & $\pm 1.9\%$ & $ - 1.5 $/$ + 1.7\% $  \\
& & Electron trigger + reco. & $-0.6$/$+0.7\%$ &  $ \pm 1.2\%$ & $ - 0.8 $/$ + 0.9\% $  \\
& & Hadronic trigger & $-1.3$/$+1.2\%$ & $\pm 1.5\%$& $\pm 3.0\%$ \\
& & Pileup & $-1.0$/$+0.9\%$ & $ \pm 0.4\%$ & $ - 0.3 $/$ + 0.2\% $  \\
& & $\MET$ modelling & $-0.0$/$+0.2\%$ & $\pm 0.2\%$ & $ \pm 0.5 \% $  \\
\cline{2-6}
& \multirow{7}{*}{\rotatebox{90}{Backg. rates}}& W+jets  &   $-2.0$/$+3.0\%$ & $-3.5$/$+2.5\%$ & $\pm5.9 \%$  \\
& & \quad light flavour (u, d, s, g) &   $-0.2$/$+0.3\%$ & $ \pm 0.4\%$ &   n/a  \\
& & \quad heavy flavour  (b, c)&  $-1.9$/$+2.9\%$  & $-3.5$/$+2.5\%$ &  n/a   \\
& & $\ttbar$  & $-0.9$/$+0.8\%$ &$\pm 1.0\%$ & $ \pm 3.3\% $  \\
& & QCD, muon & $\pm0.8\%$ & $\pm 1.7\%$ &  $ \pm 0.9\% $  \\
& & QCD, electron & $\pm 0.4\%$ & $\pm 0.8\%$ & $ - 0.4 $/$ + 0.3\% $  \\
& & $s$-, tW ch., dibosons, Z+jets & $\pm 0.3\%$ & $\pm 0.6\%$ & $\pm0.5\%$ \\
\cline{2-6}
& \multicolumn{2}{l|}{Total marginalised uncertainty} & $-7.7$/$+7.9\%$ & $-7.7$/$+7.8\%$ & n/a \\
\hline
 \multirow{8}{*}{\rotatebox{90}{Not marginalised}}  & \multicolumn{1}{l}{ }&\multicolumn{1}{l|}{Luminosity} & \multicolumn{3}{c}{$\pm 2.2\%$ } \\
\cline{2-6}
& \multirow{7}{*}{\rotatebox{90}{Theor. uncert.}} & Scale, $\ttbar$ & $-3.3$/$+1.0\%$ & $\pm 0.9\%$ & $ - 4.0 $/$ + 2.1 \% $  \\
& & Scale, W+jets  & $-2.8$/$+0.3\%$ & $- 0.0$/$+3.4\%$ & n/a  \\
& & Scale, $t$-, $s$-, tW channels & $-0.4$/$+1.0\%$ &  $\pm 0.2\%$ & $ - 2.2 $/$ +2.3 \% $  \\
& & Matching, $\ttbar$ & $\pm 1.3\%$ & $\pm 0.4\%$& $ \pm 0.4 \% $  \\
& & $t$-channel generator & $\pm 4.2\%$ & $\pm 4.6\%$ & $ \pm 2.5 \% $  \\
& & PDF & $\pm 1.3\%$ &  $\pm 1.3\%$ & $\pm 2.5\%$ \\
\cline{3-6}
& & Total theor. uncertainty& $-6.3$/$+4.8\%$ & $-4.9$/$+5.9\%$&  $-5.6$/$+4.9\%$ \\
\hline
 \multicolumn{3}{l|}{ Syst. + theor. + luminosity uncert.} &  $-8.1$/$+7.8\%$ & $-8.1$/$+8.4\%$ & $\pm 10.8\%$ \\
\hline
 \multicolumn{3}{l|}{Total (stat. + syst. + theor. + lum.)} & $-10.1$/$+9.5\%$ & $ -9.4$/$+ 10.0\%$ & $\pm 13.8\%$ \\
\hline
\end{tabular}
\end{center}
\end{table}

\section{Results}
\label{sec:results}
\subsection{Individual analyses}
The $\etalj$ analysis yields the following cross section measurements for the muon and electron channels:
\begin{align*}
\sigmatch = 73.3 \pm 10.4 \text{ (stat. + syst. + lum.)} \, \pm 4.0 \text{ (theor.)}  \,\, \text{pb}  & \quad \text{(muons),} \\
\sigmatch = 61.6 \pm 13.9 \text{ (stat. + syst. + lum.)} \, \pm 3.5 \text{ (theor.)} \,\, \text{pb}  & \quad \text{(electrons).} \\
\end{align*}

The two measurements are compatible, taking into account correlated and uncorrelated uncertainties.
The uncorrelated uncertainties include the W+jets and QCD extraction procedure, lepton reconstruction and trigger efficiencies, and the hadronic part of the trigger.

Combining the muon and electron measurements gives
\begin{align*}
\sigmatch &  = 70.0\pm 6.0  \text{ (stat.)}\pm 6.5 \text{ (syst.)} \pm 3.6\text{ (theor.)}\pm 1.5  \text{ (lum.)}\,\, \text{pb.}
\end{align*}

The Bayesian inference is performed with the data samples and the 50\% quantile is calculated as the best parameter estimate for the signal
strength $\mu$. The 84\% and 16\% quantiles are quoted as upper and lower boundaries for the $1\sigma$ credible interval.

The measured single-top-quark $t$-channel production cross section in the NN analysis is
\begin{align*}
\sigmatch = 69.7 _{ -7.0}^{ +7.2} \text{ (stat. + syst. + lum.)} \pm {3.6}\text{ (theor.)} \,\, \text{pb}  & \quad \text{(muons),} \\
\sigmatch = 65.1 _{ -8.9}^{ +9.2} \text{ (stat. + syst. + lum.)} \pm {3.5}\text{ (theor.)} \,\, \text{pb}  & \quad \text{(electrons).} \\
\end{align*}
The quoted errors are the total uncertainties including marginalised, unmarginalised, and luminosity uncertainties.
The two measurements are compatible within the uncertainties, after properly taking into account the correlated contributions. The combination of the muon and electron measurements gives
\begin{align*}
\sigmatch = 68.1\pm 4.1  \text{ (stat.)}\pm 3.4 \text{ (syst.)}{}_{ -4.3}^{ +3.3}\text{ (theor.)}\pm 1.5  \text{ (lum.)}\,\, \text{pb.}
\end{align*}

The measured single-top-quark $t$-channel production cross section in the BDT analysis is
\begin{align*}
\sigmatch =  66.6_{ -6.6}^{ +7.0} \text{ (stat. + syst. + lum.)}{}_{ -3.5}^{ +6.4} \text{ (theor.)}\,\, \text{pb}  & \quad \text{(muons),}\\
\sigmatch =  66.4_{ -7.9}^{ +8.4} \text{ (stat. + syst. + lum.)}{}_{ -5.4}^{ +5.4} \text{ (theor.)}\,\, \text{pb}  & \quad \text{(electrons),}\\
\end{align*}
and their combination gives
\begin{align*}
\sigmatch & = 66.6\pm 4.0  \text{ (stat.)}\pm 3.3 \text{ (syst.)}{}_{ -3.3}^{ +3.9}\text{ (theor.)}\pm 1.5  \text{ (lum.)}\,\, \text{pb.}  \\
\end{align*}

The results of the three analyses are consistent with the SM prediction.

\subsection{Combination}
The results of the three analyses are combined using the BLUE method.
The statistical correlation between each pair of measurements is estimated by generating dedicated pseudo-experiments. The correlation is 60\%
between NN and $\etalj$, 69\% between BDT and $\etalj$, and 74\% between NN and BDT.
Correlations for the jet energy scale and resolution, b tagging, and $\MET$ modelling between
$\etalj$ and the two multivariate analyses are expected to be small. This is because the determination of the corresponding nuisance parameters, from the marginalisation
adopted in the BDT and NN analyses, is dominated by in-situ constraints from data samples independent
of those used to determine uncertainties in the $\etalj$ analysis. The assumed correlation for those uncertainties is taken to be 20\%. The correlation has, nevertheless, been varied from
0\% to 50\%, with a corresponding variation of the central value by $-0.03$\unit{pb}, and no appreciable variation has been observed for the combined uncertainty. For trigger uncertainties, the correlation between $\etalj$ and the two multivariate analyses is more difficult to ascertain.
Varying the correlations in the combination from 0\% to 100\%
results in a variation of the central value of 0.03\unit{pb}, with no appreciable variation
of the combined uncertainty.  All other uncertainties are determined mostly from the same data
samples used by the two analyses, hence 100\% correlation is assumed.

The BLUE method is applied iteratively, as previously carried out in Ref.~\cite{Chatrchyan:2011vp}. In each iteration, the absolute uncertainty is calculated by scaling the relative uncertainties given in Table~\ref{tab:syst_combined} with the combined value from the previous iteration. This is repeated until the combined value remains constant. There are no appreciable changes with respect to the non-iterative BLUE method. The 0.03\unit{pb} variation in the central value, due to changes in correlation coefficients, is added in quadrature
to the total uncertainty. However, this results in a negligible additional contribution.

The $\chi^2$ obtained by the BLUE combination of the three analyses is 0.19, corresponding to a $p$-value of 0.90. The results of the individual analyses are consistent with each other.

The combined result of the measured single-top-quark $t$-channel production cross section at $\sqrt{s}=7$\TeV is
\begin{align}
\sigmatch &= 67.2\pm 6.1 \,\, \text{pb} = 67.2\pm 3.7  \text{ (stat.)}\pm 3.0 \text{ (syst.)} \pm 3.5\text{ (theor.)}\pm 1.5  \text{ (lum.)}\,\, \text{pb}  \nonumber
\end{align}
for an assumed top-quark mass of 172.5\GeVcc.

\subsection{\texorpdfstring{$|V_{\text{tb}}|$}{V[tb]} Extraction}
The absolute value of the CKM element $|V_{\text{tb}}|$ is determined in a similar fashion to Ref.~\cite{Chatrchyan:2011vp},
assuming that $|V_{\text{td}}|$ and $|V_{\text{ts}}|$ are much smaller than $|V_{\text{tb}}|$,
as $|V_{\text{tb}}| = \sqrt{{\sigmatch}/{\sigmatch^{\text{th}}}}$,
where $\sigmatch^{\text{th}}$ is the SM prediction calculated assuming $|V_{\text{tb}}|=1$ \cite{Kidonakis:2011wy}.
If we take into account the possible presence of an anomalous Wtb coupling, this relation modifies taking into account an anomalous form factor $f_{L_{V}}$~\cite{Wtb:anom_1, Wtb:anom_2, Rizzo:1995uv}, which is not necessarily equal to $1$ in beyond-the-SM models. We determine
\begin{equation*}
|f_{L_{V}} V_{\text{tb}}| = \sqrt{\frac{\sigmatch}{\sigmatch^{\text{th}}}} =  1.020\pm 0.046\text{ (meas.)} \pm0.017\text{ (theor.)}\,,
\end{equation*}
where the first uncertainty term contains all uncertainties of the cross section measurement including theoretical ones, and the second is the uncertainty on the SM theoretical prediction. From this result,
the confidence interval for $|V_{\text{tb}}|$, assuming the constraint $|V_{\text{tb}}| \leq 1$ and $f_{L_{V}}=1$, is determined using the unified approach of Feldman and Cousins~\cite{FC} to be
\begin{equation*}
0.92 < |V_{\text{tb}}| \le 1,\,\,\,\,\,\text{at the 95\% confidence level.}
\end{equation*}

\section{Conclusions}

The cross section of $t$-channel single-top-quark production has been measured in pp collisions using 2011 data in semileptonic top-quark decay modes with improved precision compared to the previous CMS measurement.
Two approaches have been adopted. One approach has been based on a fit of the characteristic pseudorapidity distribution of the light quark recoiling against the single top quark in the $t$-channel with background determination from data. The other has been based on two multivariate discriminators, a Neural Network and Boosted Decision Trees. The multivariate analyses reduce the impact of systematic uncertainties by simultaneously analysing phase space regions with substantial $t$-channel single-top-quark contributions, and regions where they are negligible.
The results are all consistent within uncertainties. As a consequence, all three analyses have been combined with the Best Linear Unbiased Estimator method to obtain the final result.

\begin{figure}[!ht]
 \begin{center}
  \includegraphics[width=0.8\textwidth]{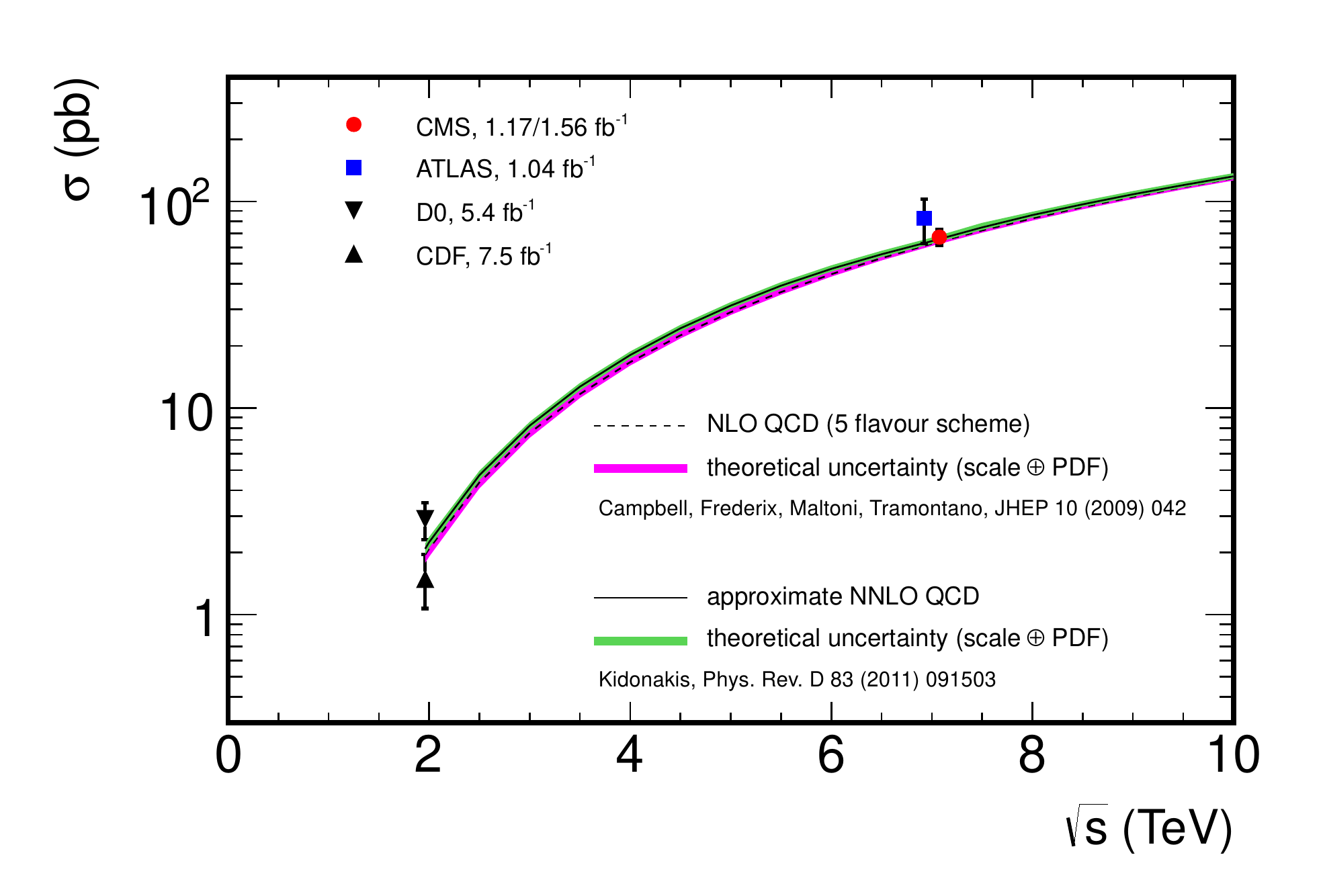}
  \caption{\label{fig:finalplot}{} The single-top-quark cross section in the $t$-channel vs. centre-of-mass energy.  The error band (width of the curve) of the SM calculation is obtained by varying the top-quark mass within its current uncertainty~\cite{Group:2009qk}, estimating the PDF uncertainty according to \textsc{hepdata} recommendations~\cite{Campbell:2006wx}, and varying the factorisation and renormalisation scales coherently by a factor of two up and down. The central values of the two SM predictions differ by 2.2\% at $\sqrt{s}=7$\TeV.}
 \end{center}
\end{figure}

The combined measurement of the single-top-quark $t$-channel cross section is $67.2\pm 6.1\unit{pb}$. This is the first measurement with a relative uncertainty below 10\%. It is in agreement with the approximate NNLO standard model prediction of $64.6^{+2.1}_{-0.7}\,^{+1.5}_{-1.7}\unit{pb}$~\cite{Kidonakis:2011wy}. Figure~\ref{fig:finalplot} compares this measurement with dedicated $t$-channel cross section measurements at the Tevatron~\cite{Abazov:2011rz,Aaltonen:2009jj}, ATLAS~\cite{Aad:2012ux}, and with the QCD expectations computed at NLO with~{\sc MCFM}~in the 5-flavour scheme~\cite{Campbell:2009gj} and at approximate NNLO~\cite{Kidonakis:2011wy}. The absolute value of the CKM matrix element $V_{\mathrm{tb}}$ is measured to be $|f_{L_{V}}V_{\mathrm{tb}}| = \sqrt{{\sigmatch}/{\sigmatch^{\mathrm{th}}}} = 1.020 \pm 0.046\text{ (meas.)} \pm 0.017\text{ (theor.)}$. Assuming $f_{L_{V}}=1$ and $|V_{\mathrm{tb}}| \leq 1$, we measure the 95\% confidence level interval $0.92 < |V_{\mathrm{tb}}| \leq 1$.

\section*{Acknowledgements}

We congratulate our colleagues in the CERN accelerator departments for the excellent performance of the LHC machine. We thank the technical and administrative staff at CERN and other CMS institutes, and acknowledge support from BMWF and FWF (Austria); FNRS and FWO (Belgium); CNPq, CAPES, FAPERJ, and FAPESP (Brazil); MES (Bulgaria); CERN; CAS, MoST, and NSFC (China); COLCIENCIAS (Colombia); MSES (Croatia); RPF (Cyprus); MEYS (Czech Republic); MoER, SF0690030s09 and ERDF (Estonia); Academy of Finland, MEC, and HIP (Finland); CEA and CNRS/IN2P3 (France); BMBF, DFG, and HGF (Germany); GSRT (Greece); OTKA and NKTH (Hungary); DAE and DST (India); IPM (Iran); SFI (Ireland); INFN (Italy); NRF and WCU (Korea); LAS (Lithuania); CINVESTAV, CONACYT, SEP, and UASLP-FAI (Mexico); MSI (New Zealand); PAEC (Pakistan); MSHE and NSC (Poland); FCT (Portugal); JINR (Armenia, Belarus, Georgia, Ukraine, Uzbekistan); MON, RosAtom, RAS and RFBR (Russia); MSTD (Serbia); SEIDI and CPAN (Spain); Swiss Funding Agencies (Switzerland); NSC (Taipei); TUBITAK and TAEK (Turkey); NASU (Ukraine); STFC (United Kingdom); DOE and NSF (USA).

Individuals have received support from the Marie-Curie programme and the European Research Council (European Union); the Leventis Foundation; the A. P. Sloan Foundation; the Alexander von Humboldt Foundation; the Austrian Science Fund (FWF); the Belgian Federal Science Policy Office; the Fonds pour la Formation \`a la Recherche dans l'Industrie et dans l'Agriculture (FRIA-Belgium); the Agentschap voor Innovatie door Wetenschap en Technologie (IWT-Belgium); the Council of Science and Industrial Research, India; the Compagnia di San Paolo (Torino); and the HOMING PLUS programme of Foundation for Polish Science, cofinanced from European Union, Regional Development Fund.

\bibliography{auto_generated}   
\cleardoublepage \appendix\section{The CMS Collaboration \label{app:collab}}\begin{sloppypar}\hyphenpenalty=5000\widowpenalty=500\clubpenalty=5000\input{TOP-11-021-authorlist.tex}\end{sloppypar}
\end{document}

%% file: TOP-11-021-authorlist.tex
\textbf{Yerevan Physics Institute,  Yerevan,  Armenia}\\*[0pt]
S.~Chatrchyan, V.~Khachatryan, A.M.~Sirunyan, A.~Tumasyan
\vskip\cmsinstskip
\textbf{Institut f\"{u}r Hochenergiephysik der OeAW,  Wien,  Austria}\\*[0pt]
W.~Adam, E.~Aguilo, T.~Bergauer, M.~Dragicevic, J.~Er\"{o}, C.~Fabjan\cmsAuthorMark{1}, M.~Friedl, R.~Fr\"{u}hwirth\cmsAuthorMark{1}, V.M.~Ghete, N.~H\"{o}rmann, J.~Hrubec, M.~Jeitler\cmsAuthorMark{1}, W.~Kiesenhofer, V.~Kn\"{u}nz, M.~Krammer\cmsAuthorMark{1}, I.~Kr\"{a}tschmer, D.~Liko, I.~Mikulec, M.~Pernicka$^{\textrm{\dag}}$, D.~Rabady\cmsAuthorMark{2}, B.~Rahbaran, C.~Rohringer, H.~Rohringer, R.~Sch\"{o}fbeck, J.~Strauss, A.~Taurok, W.~Waltenberger, C.-E.~Wulz\cmsAuthorMark{1}
% ...
\vskip\cmsinstskip
\dag:~Deceased\\
1:~~Also at Vienna University of Technology, Vienna, Austria\\
2:~~Also at CERN, European Organization for Nuclear Research, Geneva, Switzerland\\
% ...


%% file: TOP-11-021_temp.bbl
\providecommand{\href}[2]{#2}\begingroup\raggedright\begin{thebibliography}{10}%
\makeatletter
\providecommand{\hrefCMSnoop }[0]{\@secondoftwo}%
\makeatother
\providecommand{\doi}{\texttt{doi:}\begingroup \urlstyle{tt}\Url}

\bibitem{CDF-singletop}
\hrefCMSnoop {} {{ CDF} Collaboration, ``{Observation of single top quark
  production and measurement of $|V_{tb}|$ with CDF}'',} \textit{ Phys. Rev. D}
  \textbf{ 82} (2010) 112005,
  \href{http://dx.doi.org/10.1103/PhysRevD.82.112005}{\doi{10.1103/PhysRevD.82.112005}},
\href{http://www.arXiv.org/abs/1004.1181}{\texttt{ arXiv:1004.1181}}.

\bibitem{D0-singletop}
\hrefCMSnoop {} {{ D0} Collaboration, ``{Measurements of single top quark
  production cross sections and $|{V}_{tb}|$ in $p\overline{p}$ collisions at
  $\sqrt{s}=1.96\text{\,}\text{\,}\mathrm{TeV}$}'',} \textit{ Phys. Rev. D}
  \textbf{ 84} (2011) 112001,
  \href{http://dx.doi.org/10.1103/PhysRevD.84.112001}{\doi{10.1103/PhysRevD.84.112001}},
\href{http://www.arXiv.org/abs/1108.3091}{\texttt{ arXiv:1108.3091}}.

\bibitem{Group:2009qk}
\hrefCMSnoop {} {{CDF and D0 Collaborations}, ``{Combination of CDF and D0
  measurements of the single top production cross section}'',} (2009).
\href{http://www.arXiv.org/abs/0908.2171}{\texttt{ arXiv:0908.2171}}.

\bibitem{Chatrchyan:2011vp}
\hrefCMSnoop {} {{ CMS} Collaboration, ``Measurement of the $t$-channel single
  top quark production cross section in pp collisions at $\sqrt{s} =
  7$~{TeV}'',} \textit{ Phys. Rev. Lett.} \textbf{ 107} (2011) 091802,
  \href{http://dx.doi.org/10.1103/PhysRevLett.107.091802}{\doi{10.1103/PhysRevLett.107.091802}},
\href{http://www.arXiv.org/abs/1106.3052}{\texttt{ arXiv:1106.3052}}.

\bibitem{Aad:2012ux}
\hrefCMSnoop {} {{ ATLAS} Collaboration, ``{Measurement of the t-channel single
  top-quark production cross section in pp collisions at $\sqrt{s} = 7$~{TeV}
  with the ATLAS detector}'',} \textit{ Phys. Lett. B} \textbf{ 717} (2012)
  330,
  \href{http://dx.doi.org/10.1016/j.physletb.2012.09.031}{\doi{10.1016/j.physletb.2012.09.031}},
\href{http://www.arXiv.org/abs/1205.3130}{\texttt{ arXiv:1205.3130}}.

\bibitem{:2012dj}
\hrefCMSnoop {} {{ ATLAS} Collaboration, ``{Evidence for the associated
  production of a W boson and a top quark in ATLAS at $\sqrt{s} = 7~{TeV}$}'',}
  \textit{ Phys. Lett. B} \textbf{ 716} (2012) 142,
  \href{http://dx.doi.org/10.1016/j.physletb.2012.08.011}{\doi{10.1016/j.physletb.2012.08.011}},
\href{http://www.arXiv.org/abs/1205.5764}{\texttt{ arXiv:1205.5764}}.

\bibitem{Kidonakis:2011wy}
\hrefCMSnoop {} {N.~Kidonakis, ``{Next-to-next-to-leading-order collinear and
  soft gluon corrections for t-channel single top quark production}'',}
  \textit{ Phys. Rev. D} \textbf{ 83} (2011) 091503,
  \href{http://dx.doi.org/10.1103/PhysRevD.83.091503}{\doi{10.1103/PhysRevD.83.091503}},
\href{http://www.arXiv.org/abs/1103.2792}{\texttt{ arXiv:1103.2792}}.

\bibitem{BLUE}
\hrefCMSnoop {} {L.~Lyons, D.~Gibaut, and P.~Clifford, ``How to combine
  correlated estimates of a single physical quantity'',} \textit{ Nucl. Instr.
  and Meth. A} \textbf{ 270} (1988) 110,
\href{http://dx.doi.org/10.1016/0168-9002(88)90018-6}{\doi{10.1016/0168-9002(88)90018-6}}.

\bibitem{JINST}
\hrefCMSnoop {} {{ CMS} Collaboration, ``The {CMS} experiment at the {CERN}
  {LHC}'',} \textit{ JINST} \textbf{ 03} (2008) S08004,
\href{http://dx.doi.org/10.1088/1748-0221/3/08/S08004}{\doi{10.1088/1748-0221/3/08/S08004}}.

\bibitem{Re:2010bp}
\hrefCMSnoop {} {E.~Re, ``{Single-top Wt-channel production matched with parton
  showers using the POWHEG method}'',} \textit{ Eur. Phys. J. C} \textbf{ 71}
  (2011) 1547,
  \href{http://dx.doi.org/10.1140/epjc/s10052-011-1547-z}{\doi{10.1140/epjc/s10052-011-1547-z}},
\href{http://www.arXiv.org/abs/1009.2450}{\texttt{ arXiv:1009.2450}}.

\bibitem{Alioli:2010xd}
S.~Alioli\hrefCMSnoop {} { {et~al.}, ``{A general framework for implementing
  NLO calculations in shower Monte Carlo programs: the POWHEG BOX}'',} \textit{
  JHEP} \textbf{ 06} (2010) 043,
  \href{http://dx.doi.org/10.1007/JHEP06(2010)043}{\doi{10.1007/JHEP06(2010)043}},
\href{http://www.arXiv.org/abs/1002.2581}{\texttt{ arXiv:1002.2581}}.

\bibitem{Alioli:2009je}
S.~Alioli\hrefCMSnoop {} { {et~al.}, ``{NLO single-top production matched with
  shower in POWHEG: s- and t-channel contributions}'',} \textit{ JHEP} \textbf{
  09} (2009) 111,
  \href{http://dx.doi.org/10.1088/1126-6708/2009/09/111}{\doi{10.1088/1126-6708/2009/09/111}},
\href{http://www.arXiv.org/abs/0907.4076}{\texttt{ arXiv:0907.4076}}.

\bibitem{Frixione:2007vw}
\hrefCMSnoop {} {S.~Frixione, P.~Nason, and C.~Oleari, ``{Matching NLO QCD
  computations with parton shower simulations: the POWHEG method}'',} \textit{
  JHEP} \textbf{ 11} (2007) 070,
  \href{http://dx.doi.org/10.1088/1126-6708/2007/11/070}{\doi{10.1088/1126-6708/2007/11/070}},
\href{http://www.arXiv.org/abs/0709.2092}{\texttt{ arXiv:0709.2092}}.

\bibitem{pythia}
\hrefCMSnoop {} {T.~{Sj\"ostrand}, S.~Mrenna, and P.~Z. Skands, ``{PYTHIA 6.4
  physics and manual}'',} \textit{ JHEP} \textbf{ 05} (2006) 026,
  \href{http://dx.doi.org/10.1088/1126-6708/2006/05/026}{\doi{10.1088/1126-6708/2006/05/026}},
\href{http://www.arXiv.org/abs/hep-ph/0603175}{\texttt{ arXiv:hep-ph/0603175}}.

\bibitem{comphep}
E.~E. Boos\hrefCMSnoop {} { {et~al.}, ``{CompHEP - computer system for
  calculation of particle collisions at high energies }'',} Preprint 89-63/140,
  Moscow State University, Institute for Nuclear Physics, (1989).

\bibitem{madgraph}
J.~Alwall\hrefCMSnoop {} { {et~al.}, ``{MadGraph} 5: going beyond'',} \textit{
  JHEP} \textbf{ 06} (2011) 128,
  \href{http://dx.doi.org/10.1007/JHEP06(2011)128}{\doi{10.1007/JHEP06(2011)128}},
\href{http://www.arXiv.org/abs/1106.0522}{\texttt{ arXiv:1106.0522}}.

\bibitem{Alwall:2007fs}
J.~Alwall\hrefCMSnoop {} { {et~al.}, ``{Comparative study of various algorithms
  for the merging of parton showers and matrix elements in hadronic
  collisions}'',} \textit{ Eur. Phys. J. C} \textbf{ 53} (2008) 473,
  \href{http://dx.doi.org/10.1140/epjc/s10052-007-0490-5}{\doi{10.1140/epjc/s10052-007-0490-5}},
\href{http://www.arXiv.org/abs/0706.2569}{\texttt{ arXiv:0706.2569}}.

\bibitem{PDF:CTEQ6}
J.~Pumplin\hrefCMSnoop {} { {et~al.}, ``{New generation of parton distributions
  with uncertainties from global QCD analysis}'',} \textit{ JHEP} \textbf{ 07}
  (2002) 012,
  \href{http://dx.doi.org/10.1088/1126-6708/2002/07/012}{\doi{10.1088/1126-6708/2002/07/012}},
\href{http://www.arXiv.org/abs/hep-ph/0201195}{\texttt{ arXiv:hep-ph/0201195}}.

\bibitem{geant}
\hrefCMSnoop {} {S.~Agostinelli {et~al.}, ``Geant4---a simulation toolkit'',}
  \textit{ Nucl. Instrum. Meth. A} \textbf{ 506} (2003) 250,
  \href{http://dx.doi.org/10.1016/S0168-9002(03)01368-8}{\doi{10.1016/S0168-9002(03)01368-8}}.

\bibitem{btag}
\href {http://cdsweb.cern.ch/record/1427247} {{ CMS} Collaboration, ``b-Jet
  Identification in the {CMS} Experiment'',} CMS Physics Analysis Summary
  CMS-PAS-BTV-11-004, (2011).

\bibitem{top-11-014}
\hrefCMSnoop {} {{ CMS} Collaboration, ``Measurement of the charge asymmetry in
  top-quark pair production in proton-proton collisions at $\sqrt{s} =
  7\,\mathrm{TeV}$'',} \textit{ Phys. Lett. B} \textbf{ 709} (2012) 28,
  \href{http://dx.doi.org/10.1016/j.physletb.2012.01.078}{\doi{10.1016/j.physletb.2012.01.078}},
\href{http://www.arXiv.org/abs/1112.5100}{\texttt{ arXiv:1112.5100}}.

\bibitem{pflow}
\href {http://cdsweb.cern.ch/record/1194487} {{ CMS} Collaboration,
  ``Particle--Flow Event Reconstruction in {CMS} and Performance for Jets,
  Taus, and {\MET}'',} CMS Physics Analysis Summary CMS-PAS-PFT-09-001, (2009).

\bibitem{Cacciari:2008gp}
\hrefCMSnoop {} {M.~Cacciari, G.~P. Salam, and G.~Soyez, ``{The anti-$k_t$ jet
  clustering algorithm}'',} \textit{ JHEP} \textbf{ 04} (2008) 063,
  \href{http://dx.doi.org/10.1088/1126-6708/2008/04/063}{\doi{10.1088/1126-6708/2008/04/063}},
\href{http://www.arXiv.org/abs/0802.1189}{\texttt{ arXiv:0802.1189}}.

\bibitem{wbatlas}
\hrefCMSnoop {} {{ ATLAS} Collaboration, ``{Measurement of the cross section
  for the production of a W boson in association with b-jets in pp collisions
  at $\sqrt{s} = 7$~{TeV} with the ATLAS detector}'',} \textit{ Phys. Lett. B}
  \textbf{ 707} (2012) 418,
  \href{http://dx.doi.org/10.1016/j.physletb.2011.12.046}{\doi{10.1016/j.physletb.2011.12.046}},
\href{http://www.arXiv.org/abs/1109.1470}{\texttt{ arXiv:1109.1470}}.

\bibitem{Motylinski:2009kt}
\hrefCMSnoop {} {P.~Motylinski, ``{Angular correlations in t-channel single top
  production at the LHC}'',} \textit{ Phys. Rev. D} \textbf{ 80} (2009) 074015,
  \href{http://dx.doi.org/10.1103/PhysRevD.80.074015}{\doi{10.1103/PhysRevD.80.074015}},
\href{http://www.arXiv.org/abs/0905.4754}{\texttt{ arXiv:0905.4754}}.

\bibitem{Feindt:2004}
\hrefCMSnoop {} {M.~Feindt, ``A neural bayesian estimator for conditional
  probability densities'',} (2004).
  \href{http://www.arXiv.org/abs/physics/0402093}{\texttt{
  arXiv:physics/0402093}}.

\bibitem{Feindt:2006pm}
\hrefCMSnoop {} {M.~Feindt and U.~Kerzel, ``{The NeuroBayes neural network
  package}'',} \textit{ Nucl. Instrum. Meth. A} \textbf{ 559} (2006) 190,
\href{http://dx.doi.org/10.1016/j.nima.2005.11.166}{\doi{10.1016/j.nima.2005.11.166}}.

\bibitem{efron93bootstrap}
B.~Efron and R.~J. Tibshirani, ``{An Introduction to the Bootstrap}''.
\newblock Chapman \& Hall, New York, 1993.

\bibitem{BDT:Vispa}
H.-P. Bretz\hrefCMSnoop {} { {et~al.}, ``A development environment for visual
  physics analysis'',} \textit{ JINST} \textbf{ 07} (2012) T08005,
  \href{http://dx.doi.org/10.1088/1748-0221/7/08/T08005}{\doi{10.1088/1748-0221/7/08/T08005}},
  \href{http://www.arXiv.org/abs/1205.4912}{\texttt{ arXiv:1205.4912}}.

\bibitem{tmva}
A.~Hoecker\hrefCMSnoop {} { {et~al.}, ``{TMVA: Toolkit for Multivariate Data
  Analysis}'',} (2007).
\href{http://www.arXiv.org/abs/physics/0703039}{\texttt{
  arXiv:physics/0703039}}.

\bibitem{Jaynes2003}
E.~T. Jaynes, ``Probability Theory: The Logic of Science''.
\newblock Cambridge University Press, Cambridge, 2003.

\bibitem{theta}
\href {http://www-ekp.physik.uni-karlsruhe.de/~ott/theta/files/theta.pdf}
  {T.~{M\"uller}, J.~Ott, and J.~Wagner-Kuhr, ``\textsc{Theta} -- a framework
  for template-based statistical modeling and inference'',} Preprint
  CMS/2012-1, (2012).

\bibitem{CMS-PAS-JME-10-010}
\hrefCMSnoop {} {{ CMS} Collaboration, ``Determination of jet energy
  calibration and transverse momentum resolution in CMS'',} \textit{ JINST}
  \textbf{ 06} (2011) 11002,
  \href{http://dx.doi.org/10.1088/1748-0221/6/11/P11002}{\doi{10.1088/1748-0221/6/11/P11002}},
  \href{http://www.arXiv.org/abs/1107.4277}{\texttt{ arXiv:1107.4277}}.

\bibitem{CMS-PAS-JME-10-014}
\href {http://cdsweb.cern.ch/record/1339945} {{ CMS} Collaboration, ``Jet
  Energy Resolution in CMS at $\sqrt{s}=7$ {TeV}'',} CMS Physics Analysis
  Summary CMS-PAS-JME-10-014, (2010).

\bibitem{Khachatryan:2010xn}
\hrefCMSnoop {} {{ CMS} Collaboration, ``{Measurements of inclusive W and Z
  cross sections in pp collisions at $\sqrt{s} = 7$~{TeV}}'',} \textit{ JHEP}
  \textbf{ 01} (2011) 080,
  \href{http://dx.doi.org/10.1007/JHEP01(2011)080}{\doi{10.1007/JHEP01(2011)080}},
\href{http://www.arXiv.org/abs/1012.2466}{\texttt{ arXiv:1012.2466}}.

\bibitem{lumi}
\href {http://cdsweb.cern.ch/record/1434360} {{ CMS} Collaboration, ``Absolute
  Calibration of the Luminosity Measurement at {CMS}: {W}inter 2012 Update'',}
  CMS Physics Analysis Summary CMS-PAS-SMP-12-008, (2012).

\bibitem{top-10-003}
\hrefCMSnoop {} {{ CMS} Collaboration, ``Measurement of the \ttbar production
  cross section in {$\Pp\Pp$} collisions at 7 {TeV} in lepton + jets events
  using {$\cPqb$}-quark jet identification'',} \textit{ Phys. Rev. D} \textbf{
  84} (2011) 092004,
  \href{http://dx.doi.org/10.1103/PhysRevD.84.092004}{\doi{10.1103/PhysRevD.84.092004}},
\href{http://www.arXiv.org/abs/1108.3773}{\texttt{ arXiv:1108.3773}}.

\bibitem{barlow_beeston}
\hrefCMSnoop {} {R.~J. Barlow and C.~Beeston, ``{Fitting using finite Monte
  Carlo samples}'',} \textit{ Comput. Phys. Commun.} \textbf{ 77} (1993) 219,
\href{http://dx.doi.org/10.1016/0010-4655(93)90005-W}{\doi{10.1016/0010-4655(93)90005-W}}.

\bibitem{bb_light}
\href {http://cdsweb.cern.ch/record/1306523} {J.~S. Conway, ``Nuisance
  parameters in likelihoods for multisource spectra'',} in \textit{
  {Proceedings of {PHYSTAT} 2011 Workshop on Statistical Issues Related to
  Discovery Claims in Search Experiments and Unfolding}}, H.~B. Prosper and
  L.~Lyons, eds., p.~115.
\newblock CERN, 2011.
\newblock \href{http://www.arXiv.org/abs/1103.0354}{\texttt{ arXiv:1103.0354}}.

\bibitem{mlm}
M.~L. Mangano\hrefCMSnoop {} { {et~al.}, ``{Matching matrix elements and shower
  evolution for top-quark production in hadronic collisions}'',} \textit{ JHEP}
  \textbf{ 01} (2007) 013,
  \href{http://dx.doi.org/10.1088/1126-6708/2007/01/013}{\doi{10.1088/1126-6708/2007/01/013}},
\href{http://www.arXiv.org/abs/hep-ph/0611129}{\texttt{ arXiv:hep-ph/0611129}}.

\bibitem{Campbell:2009ss}
J.~M. Campbell\hrefCMSnoop {} { {et~al.}, ``{Next-to-leading-order predictions
  for t-channel single-top production at hadron colliders}'',} \textit{ Phys.
  Rev. Lett.} \textbf{ 102} (2009) 182003,
  \href{http://dx.doi.org/10.1103/PhysRevLett.102.182003}{\doi{10.1103/PhysRevLett.102.182003}},
\href{http://www.arXiv.org/abs/0903.0005}{\texttt{ arXiv:0903.0005}}.

\bibitem{Frederix:2012dh}
\hrefCMSnoop {} {R.~Frederix, E.~Re, and P.~Torrielli, ``{Single-top t-channel
  hadroproduction in the four-flavour scheme with POWHEG and aMC@NLO}'',}
  (2012).
\href{http://www.arXiv.org/abs/1207.5391}{\texttt{ arXiv:1207.5391}}.

\bibitem{PDF:CTEQ10}
H.-L. Lai\hrefCMSnoop {} { {et~al.}, ``{New parton distributions for collider
  physics}'',} \textit{ Phys. Rev. D} \textbf{ 82} (2010) 074024,
  \href{http://dx.doi.org/10.1103/PhysRevD.82.074024}{\doi{10.1103/PhysRevD.82.074024}},
\href{http://www.arXiv.org/abs/1007.2241}{\texttt{ arXiv:1007.2241}}.

\bibitem{PDF:LHAPDF}
\hrefCMSnoop {} {M.~R. Whalley, D.~Bourilkov, and R.~C. Group, ``{The Les
  Houches accord PDFs (LHAPDF) and LHAGLUE}'',} (2005).
\href{http://www.arXiv.org/abs/hep-ph/0508110}{\texttt{ arXiv:hep-ph/0508110}}.

\bibitem{Wtb:anom_1}
\hrefCMSnoop {} {J.~A. Aguilar-Saavedra, ``{A minimal set of top anomalous
  couplings}'',} \textit{ Nucl. Phys. B} \textbf{ 812} (2009) 181,
  \href{http://dx.doi.org/10.1016/j.nuclphysb.2008.12.012}{\doi{10.1016/j.nuclphysb.2008.12.012}},
\href{http://www.arXiv.org/abs/0811.3842}{\texttt{ arXiv:0811.3842}}.

\bibitem{Wtb:anom_2}
\hrefCMSnoop {} {G.~L. Kane, G.~A. Ladinsky, and C.~P. Yuan, ``{Using the top
  quark for testing standard model polarization and CP predictions}'',}
  \textit{ Phys. Rev. D} \textbf{ 45} (1992) 124,
\href{http://dx.doi.org/10.1103/PhysRevD.45.124}{\doi{10.1103/PhysRevD.45.124}}.

\bibitem{Rizzo:1995uv}
\hrefCMSnoop {} {T.~G. Rizzo, ``{Single top quark production as a probe for
  anomalous moments at hadron colliders}'',} \textit{ Phys. Rev. D} \textbf{
  53} (1996) 6218,
  \href{http://dx.doi.org/10.1103/PhysRevD.53.6218}{\doi{10.1103/PhysRevD.53.6218}},
\href{http://www.arXiv.org/abs/hep-ph/9506351}{\texttt{ arXiv:hep-ph/9506351}}.

\bibitem{FC}
\hrefCMSnoop {} {G.~J. Feldman and R.~D. Cousins, ``A unified approach to the
  classical statistical analysis of small signals'',} \textit{ Phys. Rev. D}
  \textbf{ 57} (1998) 3873,
  \href{http://dx.doi.org/10.1103/PhysRevD.57.3873}{\doi{10.1103/PhysRevD.57.3873}},
\href{http://www.arXiv.org/abs/physics/9711021}{\texttt{
  arXiv:physics/9711021}}.

\bibitem{Abazov:2011rz}
\hrefCMSnoop {} {{ D0} Collaboration, ``{Model-independent measurement of
  $t$-channel single top quark production in $p\bar{p}$ collisions at
  $\sqrt{s}=1.96$ TeV}'',} \textit{ Phys. Lett. B} \textbf{ 705} (2011) 313,
  \href{http://dx.doi.org/10.1016/j.physletb.2011.10.035}{\doi{10.1016/j.physletb.2011.10.035}},
\href{http://www.arXiv.org/abs/1105.2788}{\texttt{ arXiv:1105.2788}}.

\bibitem{Aaltonen:2009jj}
\hrefCMSnoop {} {{ CDF} Collaboration, ``{First observation of electroweak
  single top quark production}'',} \textit{ Phys. Rev. Lett.} \textbf{ 103}
  (2009) 092002,
  \href{http://dx.doi.org/10.1103/PhysRevLett.103.092002}{\doi{10.1103/PhysRevLett.103.092002}},
\href{http://www.arXiv.org/abs/0903.0885}{\texttt{ arXiv:0903.0885}}.

\bibitem{Campbell:2009gj}
J.~M. Campbell\hrefCMSnoop {} { {et~al.}, ``{NLO predictions for t-channel
  production of single top and fourth generation quarks at hadron
  colliders}'',} \textit{ JHEP} \textbf{ 10} (2009) 042,
  \href{http://dx.doi.org/10.1088/1126-6708/2009/10/042}{\doi{10.1088/1126-6708/2009/10/042}},
\href{http://www.arXiv.org/abs/0907.3933}{\texttt{ arXiv:0907.3933}}.

\bibitem{Campbell:2006wx}
\hrefCMSnoop {} {J.~M. Campbell, J.~W. Huston, and W.~J. Stirling, ``{Hard
  interactions of quarks and gluons: A primer for LHC physics}'',} \textit{
  Rept. Prog. Phys.} \textbf{ 70} (2007) 89,
  \href{http://dx.doi.org/10.1088/0034-4885/70/1/R02}{\doi{10.1088/0034-4885/70/1/R02}},
\href{http://www.arXiv.org/abs/hep-ph/0611148}{\texttt{ arXiv:hep-ph/0611148}}.

\end{thebibliography}\endgroup
